\documentclass{aa}
\usepackage{color}
\usepackage{amsmath}
\usepackage{float}
\usepackage{natbib}
\usepackage{graphicx}
\usepackage{graphicx}
\usepackage{ulem}
\usepackage[varg]{txfonts}
\usepackage{lscape}
\usepackage{xspace}

\definecolor{magenta}{rgb}{1.0, 0.0, 1.0}
\definecolor{mygray}{gray}{0.6}

\def\mearth{$\,{\rm M}_{\oplus}$\xspace}

\def\h2o{H$_2$O}

\def\kgm3{kg\,m$^{-3}$}
\def\Jkg3{J\,kg$^{-3}$}

\defcitealias{Brouwers2018}{BVO18}

\begin{document} 

    \title{How planets grow by pebble accretion}
   \subtitle{II: Analytical calculations on the evolution of polluted envelopes}
   
   \author{M. G. Brouwers \inst{1,3}
          \and
          C. W. Ormel \inst{2,3}
          }

   \institute{Institute of Astronomy, University of Cambridge, Madingley Road, Cambridge CB3 0HA
   \and Department of Astronomy, Tsinghua University, Haidian DS 100084, Beijing, China
   \and Anton Pannekoek Institute, University of Amsterdam, Science Park 904, PO box 94249, Amsterdam, The Netherlands
   \\
              \email{mgb52@cam.ac.uk; chrisormel@tsinghua.edu.cn}
             }

  \abstract
{Proto-planets embedded in their natal disks acquire hot envelopes as they grow and accrete solids. This ensures that the material they accrete -- pebbles, as well as (small) planetesimals -- will vaporize to enrich their atmospheres. Enrichment modifies an envelope's structure and significantly alters its further evolution.}
{Our aim is to describe the formation of planets with polluted envelopes from the moment that impactors begin to sublimate to beyond the disk's eventual dissipation.}
{We constructed an analytical interior structure model, characterized by a hot and uniformly mixed high-Z vapor layer surrounding the core, located below the usual unpolluted radiative-convective regions. Our model assumes an ideal equation of state and focuses on identifying trends rather than precise calculations. The expressions we derived are applicable to all single-species pollutants, but we used SiO$_2$ to visualize our results.}
{The evolution of planets with uniformly mixed polluted envelopes follows four potential phases. Initially, the central core grows directly through impacts and rainout until the envelope becomes hot enough to vaporize and absorb all incoming solids. We find that a planet reaches runaway accretion when the sum of its core and vapor mass exceeds a value that we refer to as the critical metal mass -- a criterion that supersedes the traditional critical core mass. The critical metal mass scales positively with both the pollutant's evaporation temperature and with the planet's core mass. Hence, planets at shorter orbital separations require the accretion of more solids to reach runaway as they accrete less volatile materials. If the solids accretion rate dries up, we identify the decline of the mean molecular weight - dilution - as a mechanism to limit gas accretion during a polluted planet's embedded cooling phase. When the disk ultimately dissipates, the envelope's inner temperature declines and its vapor eventually rains out, augmenting the mass of the core. The energy release that accompanies this does not result in significant mass-loss, as it only occurs after the planet has substantially contracted.}
{}

   
   \maketitle
\begin{figure*}[t!] 
\centering
\includegraphics[width=\hsize]{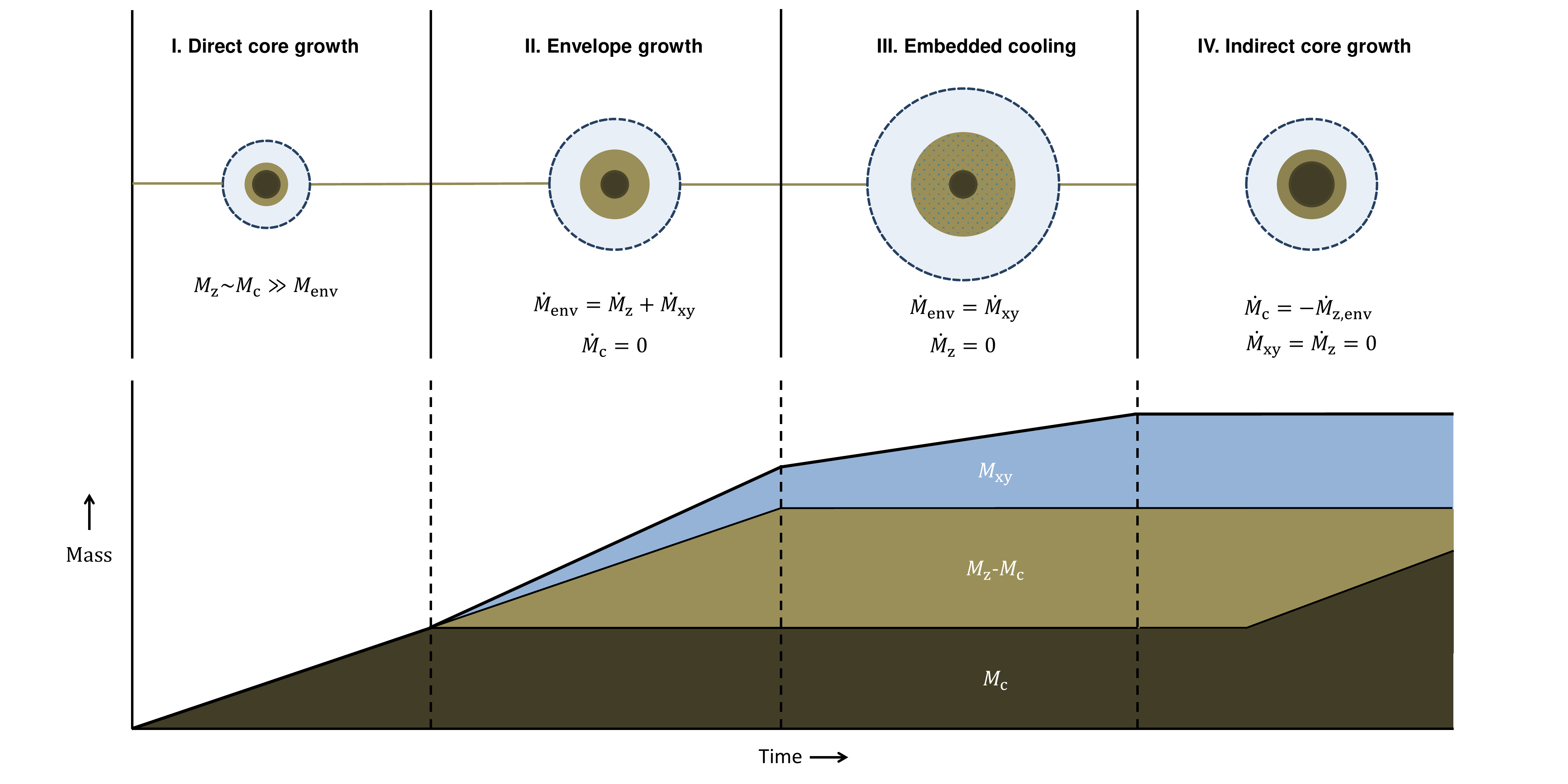} 
\caption{Sketch of the four evolutionary phases of polluted envelopes (excluding photo-evaporation). During the first phase of direct core growth, a portion of the high-Z material makes it through to the core. This contrasts with the second phase, where only the envelope ($M_\mathrm{env} = M_\mathrm{xy} + M_\mathrm{z} - M_\mathrm{c}$) gains additional mass and all accreted solids are absorbed as vapor ($M_\mathrm{z} - M_\mathrm{c}$). Phase II could terminate when the planet enters runaway accretion. Otherwise, if the solids mass flux dries up while the disk is still present, the planet enters a Kelvin-Helmholtz contraction phase (III). This leads to further gas accretion and to the dilution of the envelope. When the disk ultimately dissipates, there is no gas left to accrete and the envelope contracts (IV). This eventually leads to the (partial) sedimentation of high-Z vapor and to the indirect growth of the core. \label{fig:sketch_phases}} 
\end{figure*}
%

\section{Introduction}
The collective observational effort over the last decade has revealed that exoplanets are very abundant and far more diverse than initially anticipated. Surveys predict planetary occurrence rates around unity, at least around the most well-studied population of main-sequence dwarfs \citep{Borucki2011, Berta2015, Dressing2015, Winn2015, Mulders2019}. From these observations, studies generally discern three categories, depending on the planet's size and composition. At the lower mass range terrestrial planets like Mars and Earth reside, too small to maintain more than a tenuous atmosphere, which is negligible in mass compared to the core. At the other extreme we find gas giants like Jupiter and Saturn. These are interpreted in the standard core-accretion scenario as planets whose cores managed to grow to a sufficient size during the disk lifetime to undergo rapid gas accretion \citep{Mizuno1980,Pollack1996,Lissauer2007}. Yet by far the most abundant planets appear to be those with intermediate masses, roughly between that of Earth and Neptune, typically located at short-period orbits \citep{Fressin2013, Howard2013, Silburt2015, Mulders2019}. A portion of these have densities consistent with pure rock and are referred to as super-Earths. Some studies suggest that they had their atmospheres stripped by photo-evaporation \citep[e.g.,][]{Lopez2013, OwenWu2017, Jin2018} or through the release of heat from their cores \citep{Ginzburg2018, Gupta2018}. The majority of close-in, intermediate-mass planets are substantially more voluminous, however, and must contain envelopes that carry at least a few percent of their mass \citep{Lopez2012, Lopez2014, Marcy2014a, Rogers2015, Lozovsky2018}, in many instances more than can be produced just by outgassing \citep{Rogers2010a}. These planets are typically referred to as Sub- or Mini-Neptunes.

Explaining how the intermediate-mass planets formed is the subject of much recent and ongoing work (e.g., \citealt{Dawson2015, Lee2016, Alessi2017}; \citealt{Venturini2017, Bodenheimer2018}; \citealt{Raymond2018, Bitsch2019, Lambrechts2019}) . In order to appear as we observe them, they require sufficient time and mass to attract their gaseous envelopes before the disk dissipates, but must also not accrete so fast that they become gas giants. This problem is twofold, as halting solids accretion prematurely accelerates their cooling and the onset of runaway accretion \citep{LeeEtal2014}. Within current formation models, the proposed solutions to this problem are either to assume very opaque envelopes owing to the hypothesized presence of small grains or to consider the dynamical coupling between the planetary envelope and disk that can result in the continual recycling of envelope gas \citep{Ormel2015, Fung2015, Alibert2017, Cimerman2017}.

The alternative is to conceptually expand on the core accretion model. One of the most natural ways to do this is to account for the sublimation of impactors on their way to the core. The likelihood of their mass-loss is especially obvious in the scenario of pebble accretion \citep{Ormel2010, Lambrechts2012}, where a planet accretes small mm- to m-sized pebbles instead of km-sized planetesimals. Such small objects quickly ablate when they encounter gas with temperatures in excess of their evaporation temperature, a condition that is quickly reached in the hot primordial envelopes (\citealt{Alibert2017};  \citealt[][hereafter BVO18]{Brouwers2018}). But even km-sized planetesimals are vulnerable and are expected to break up when they impact the envelopes of super-Earth sized planets (\citealt{Podolak1988, Mcauliffe2006, Mordasini2015, Pinhas2016}; \citetalias{Brouwers2018}; \citealt{Valletta2018}).
As a result, these envelopes become polluted with high-Z vapor and their interiors consequently become significantly more hot and dense \citep{Iaroslavitz2007, Lozovsky2017, Bodenheimer2018}. One of the most prominent effects of envelope pollution is that the planet sucks in significantly more nebular gas if compositional mixing proceeds efficiently, triggering runaway accretion at a lower mass \citep{Stevenson1982, Wuchterl1993, Venturini2015, Venturini2016}.

Previous works on envelope pollution have been numerically driven and mostly focus on a planet's accretion phase. We instead construct the first analytical structure model, characterized by a perfectly mixed high-Z vapor layer, located below the usual radiative-convective envelope \citep[e.g.,][]{Inamdar2015, Lee2015, Piso2015, Ginzburg2016}. The downside of our analytical approach is that it necessitates for us to make significant simplifications, including the use of an ideal equation of state, therefore restricting the accuracy of our quantitative estimations. The advantage is that it allows us to reveal the physical cause behind emerging trends and to analyze the complete internal evolution of a polluted planet, including the stages of post-accretion cooling. As such, our research complements previous numerical works and identifies areas where their follow-up is especially pertinent.

We find that the effects of envelope pollution range far beyond the influence on a planet's critical mass, and that the presence of vapor with a high mean molecular weight fundamentally changes a planet's evolution in a number of ways. Figure \ref{fig:sketch_phases} sketches the four potential phases that a polluted envelope goes through. Initially, during the first phase, the majority of the accreted solids make it to the core. But as its internal temperature rises, the envelope absorbs increasing amounts of vapor until the core stops growing entirely. We refer to the subsequent evolutionary period of pure envelope growth as phase II. This contrasts with the classical evolution without pollution, where the first two phases are indistinguishable. Unless the planet accretes sufficient material to become a gas giant, its accretion of solids ultimately ends and the planet enters a phase of cooling. Embedded cooling (III) results in the accretion of nebular gas and the compositional dilution of the interior. In contrast, cooling after the dispersal of the disk gas (IV) is often associated with mass-loss in the form of UV-radiation and vacuum-induced outflows \citep{Lopez2013, OwenWu2017, Jin2018, Ginzburg2018, Gupta2018}. We suggest that this final phase also proceeds in a fundamentally different manner for planets with polluted envelopes, as their cores can enter a second phase of (indirect) growth due to the sedimentation of solids from the super-saturated envelope, without the occurrence of significant mass-loss.

This paper is organized as follows. We lay out the general structure of our interior model in Sect. \ref{sect:interior_structure}. Sect. \ref{sect:phase_I} begins our evolutionary description of polluted envelopes with an analysis of core growth through impacts and rainout. We then take the evolution one step further in Sect. \ref{sect:phase_II}, where we describe the subsequent phase of polluted envelope growth and derive an approximate expression for their runaway criterion. Finally, we describe in Sect. \ref{sect:phase_III} how their cooling proceeds when the primordial disk is still present, and in Sect. \ref{sect:phase_IV} what happens after it has dissipated. We discuss the implications of our findings in Sect. \ref{sect:discussion} and our conclusions in Sect. \ref{sect:conclusion}.

\section{Interior structure model}\label{sect:interior_structure}
\begin{table*}
\caption{Default values for envelope, impactor, and disk parameters.}
\label{table:default_parameters}
\centering
\begin{tabular*}{0.9\textwidth}{l l l l}
\hline\hline  
Symbol &  Description & Value & Comments \\    
\hline    
$\gamma_\mathrm{xy}$   & Adiabatic index intermediate region  & 1.45 &   \\         
$\gamma_\mathrm{z}$, $\gamma_\mathrm{g}$   & Adiabatic index high-Z region (pure, mixed)  & 1.2, 1.25 & (a)  \\
$\mu_\mathrm{xy}$   & Mean molecular weight nebular gas  & 2.34 u &   \\         
$\mu_\mathrm{z}$   & Mean molecular weight silicate vapor  & 60 u &  \\
$\rho_\mathrm{i} = \rho_\mathrm{c}$   & Density impactors and core  & $3.2 \; \mathrm{g} \; \mathrm{cm}^{-3}$ & (b) \\
$u_\mathrm{evap}$   & Latent heat silicates  & $1.5 \cdot 10^{-11} \; \mathrm{erg} \; \mathrm{g}^{-1}$ &  (c) \\
$T_\mathrm{vap}$   & Outer boundary temperature high-Z region  & 2500 K &  \\
$\dot{M}_\mathrm{z}$   & Pebble accretion rate  & $10^{-5}$ \mearth{} $\mathrm{yr}^{-1}$ &  (d) \\
$\kappa_\mathrm{cst} $   & Opacity prefactor  & $10^{-8}\; \mathrm{cm}^{4} \; \mathrm{g}^{-\frac{5}{3}} \; \mathrm{K}^{-3}$ &  (e) \\
$\beta$, $\delta$   & Opacity power-law scaling with density, temperature  & $\frac{2}{3}, 3$ &  (e) \\
$T_\mathrm{disk,5.2}$, $\rho_\mathrm{disk,5.2} $   & Disk temperature and density at 5.2 AU  & 150 K, $5 \cdot 10^{-11} \; \mathrm{g} \; \mathrm{cm}^{-3}$  &  (f) \\
\hline\hline                                    
\end{tabular*}
\tablefoot{\tablefoottext{a}{We choose the mixed index slightly higher to represent the combined effects of compositional mixing, potential dissociation and ionization.}\; \tablefoottext{b}{Assuming no core compression.}\; \tablefoottext{c}{From \citet{Melosh2008}.}\; \tablefoottext{d}{Typical accretion rate from} \citet{Lambrechts2014}.\; \tablefoottext{e}{Molecular opacity from \citet{Bell1994}.}\;  \tablefoottext{f}{Minimum Mass Solar Nebula; $T_\mathrm{disk} \propto d^{-\frac{1}{2}}$ and $\rho_\mathrm{disk} \propto d^{-\frac{11}{4}}$ \citep{Weidenschilling1977, Hayashi1981}}.}
\end{table*}
Before we turn to planetary evolution, we first outline the necessary structure equations. For simplicity and convenience, we assume that growing planets remain fully embedded in a Minimum Mass Solar Nebula until the disk dissipates \citep[e.g.,][]{Weidenschilling1977, Hayashi1981}, although we note that the Kepler planets may have formed in more massive nebulae \citep{Chiang2013, Schlichting_2014}. Like any static model, we neglect the hydrodynamic term and assume that embedded envelopes are continually stabilized by the inflow of nebular gas. Their interior is then described by the well known hydrostatic structure equations \citep[e.g.,][]{Kippenhahn1989, Benacquista2013}:
\begin{subequations}
\begin{align}
    \frac{\partial r}{\partial m} &= \frac{1}{4\pi r^2 \rho_\mathrm{g}}, \label{eq:dr_dm} \\
    \frac{\partial P_\mathrm{g}}{\partial m} &= -\frac{G m}{4\pi r^4}, \label{eq:dP_dm} \\
    \frac{\partial T}{\partial m} &= \frac{\partial P_\mathrm{g}}{\partial m} \frac{T_\mathrm{g}}{P_\mathrm{g}} \; \mathrm{min}\big(\nabla_\mathrm{rad}, \nabla_\mathrm{conv}\big), \label{eq:dT_dm}
\end{align}
\end{subequations}
where $r$ and $m$ are the planet's local radius and interior mass and $P_\mathrm{g}, T_\mathrm{g}, \rho_\mathrm{g}$ are the envelope's local gas pressure, temperature, and density

\begin{figure}[t!] 
\centering
\includegraphics[width=\hsize]{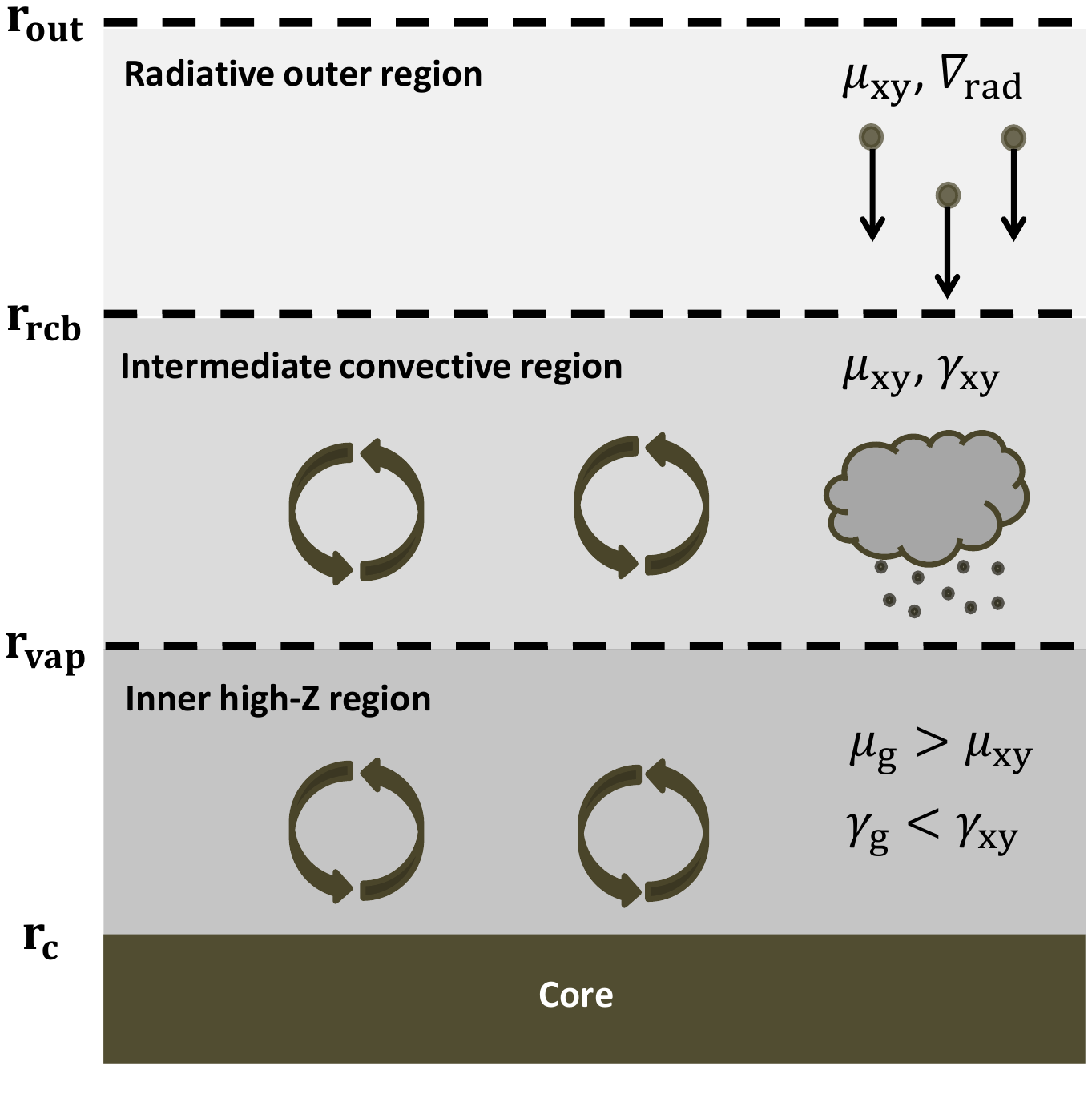} 
\caption{Sketch of the three thermodynamic zones of an envelope during its formation in a gaseous disk. Radiation dominates heat transport outside the radiative-convective boundary, while the interior is convectively unstable. The two outer regions are composed of nebular gas. In contrast, the inner high-Z region is formed by vapor from impactors and consists entirely of high-Z material until it becomes sufficiently large to mix with significant fractions of hydrogen and helium from the outer layers (see Sect. \ref{sect:uniform_mixing}). We simplify the steep compositional gradient to a step-function at $r_\mathrm{vap}$ with a continuous temperature. \label{fig:zones}} 
\end{figure}

While a planet is embedded, its structure equations end in the disk conditions at the outer boundary ($r_\mathrm{out}$). While some works take a reduced value to incorporate the thermodynamic effects of envelope recycling \citep{Lissauer2009, Dangelo2013}, we stick to the traditional convention for simplicity and identify the outer edge of an embedded planet as the lesser between the Hill ($r_\mathrm{H}$) and the Bondi ($r_\mathrm{B}$) radii \citep{Pollack1996, Benvenuto2005, Hubickyj2005, Lee2015}:
\begin{equation}
    r_\mathrm{B} = \frac{G \mu_\mathrm{xy} M_\mathrm{p}}{k_\mathrm{B} T_\mathrm{disk}} \;,\; 
    r_\mathrm{H} = d {\bigg(\frac{M_\mathrm{p}}{3 M_\mathrm{\star}}\bigg)}^\frac{1}{3},
\end{equation}
where $M_\mathrm{p}$ and $M_\mathrm{\star}$ are the planetary and stellar mass, $d$ is their separation and $T_\mathrm{disk}$, $\mu_\mathrm{xy}$ are the disk's temperature and mean molecular weight.

Moving deeper into the envelope, Eq. \ref{eq:dT_dm} distinguishes between the regions where energy transport is modulated by either radiation or convection \citep[e.g.,][]{Kippenhahn1989, Benacquista2013}:
\begin{equation}\label{eq:nabla_rad}
    \nabla_\mathrm{rad} = \frac{3 \kappa L P_\mathrm{g}}{64 \pi \sigma G m T_\mathrm{g}^4} \;,\;  \nabla_\mathrm{conv} \approx \nabla_\mathrm{ad} = \frac{\gamma -1}{\gamma},
\end{equation}
where $\kappa$ and $L$ are the local Rosseland-mean opacity and luminosity, $\gamma$ is the local adiabatic index and $\sigma$, $G$ are the Stefan-Boltzmann and gravitational constants. Radiation typically dominates towards the planet's outer edge unless this region is abundant in small grains. The energy transport at greater depths, where the envelope is unstable due to the Schwarschild criterion, is instead modulated by convection. We do not account for compositional gradients with the Ledoux criterion in our model (see Sect. \ref{sect:high-Z region}).

Numerical studies have shown the key difference between polluted and metal-free envelopes to be the presence of high-Z vapor in the envelope (\citealt{Iaroslavitz2007, Lozovsky2017, Venturini2015, Venturini2016, Bodenheimer2018}; \citetalias{Brouwers2018}). Depending on the volatility of the accreted solids and the planet's location in the disk, this pollution can affect the entire envelope (as in \citet{Venturini2015, Venturini2016} with water vapor), or be limited to the hot region around the core (as in \citet{Iaroslavitz2007, Lozovsky2017, Bodenheimer2018} with silicates). In our analytical model, we account for pollution by modifying the composition interior to a radius $r_\mathrm{vap}$ where the temperature is sufficient for vapor to exist. The result is a three-layer envelope with a uniform composition in each zone, qualitatively sketched in Fig. \ref{fig:zones}.

Ideally, the gas density in our model should be provided by a real mixed-composition equation of state \citep[e.g.,][]{Saumon1995, Faik2018} and account for dissociation, ionization and chemical reactions. Such effects are especially important in polluted envelopes as they reach can much higher internal temperatures than their metal-poor counterparts. However, our analytical approach necessitates the use of ideal gases and we thus cannot capture this complexity. An ideal gas is characterized by a high degree of compressibility and typically leads to an overestimation of the density. This significantly restricts the accuracy of our model to order-of-magnitude predictions and provides our most important limitation.

\subsection{Radiative outer region}\label{sect:radiative region}
Planets that form in the outer disk or those whose envelopes contain few grains (i.e. due to their coagulation and settling), typically contain an outer radiative region where the energy is transported by radiation. In other cases, such a region forms when the planet cools down and becomes less luminous. Its relative remoteness from the core and its disk-like conditions imply a small temperature gradient in Eq. \ref{eq:nabla_rad}. We use this to simplify the radiative part of the envelope as isothermal. We have explicitly checked the validity of this assumption by comparing it to a first-order Taylor approximation of the temperature gradient as done by \citet{Ormel2012} and found a typical resulting temperature difference at the radiative-convective boundary of only a few percent for grain-free envelopes. Integrating Eq. \ref{eq:dP_dm} using Eq. \ref{eq:dr_dm} with the ideal gas law yields the steep pressure and density curves in this radiative regime:
\begin{subequations}
\begin{align}
    \rho(r) = \rho_\mathrm{disk} \mathrm{exp}\left({\frac{r_\mathrm{B}}{r} - \frac{r_\mathrm{B}}{r_\mathrm{out}}}\right), \label{eq:rho_rad} \\
    P(r) = P_\mathrm{disk} \mathrm{exp}\left({\frac{r_\mathrm{B}}{r} - \frac{r_\mathrm{B}}{r_\mathrm{out}}}\right), \label{eq:P_rad}
\end{align}
\end{subequations}
where $\rho_\mathrm{disk}, P_\mathrm{disk}$ are the local disk density and pressure. These relations are valid down to the radiative-convective boundary $r_\mathrm{rcb}$, where the atmosphere becomes convectively unstable according to the Schwarschild criterion. The location of this transition point is a function of its local gas density $\rho_\mathrm{rcb}$ and follows directly from Eq. \ref{eq:rho_rad}:
\begin{equation}
    r_\mathrm{rcb} = \frac{r_\mathrm{B}}{\mathrm{log}(\rho_\mathrm{rcb}/\rho_\mathrm{disk}) + r_\mathrm{B}/r_\mathrm{out}}. \label{eq:r_rcb}
\end{equation}
In order to determine $\rho_\mathrm{rcb}$, we need to approximate the luminosity and opacity. For the latter, we take a general power-law expression of the form:
\begin{equation}\label{eq:opacity_law}
    \kappa_\mathrm{g} = \kappa_\mathrm{cst} \rho_\mathrm{g}^\beta T_\mathrm{g}^\delta  \; \mathrm{cm}^2 \mathrm{g}^{-1}.
\end{equation}
In our nominal case, we assume that there are no grains present at the radiative-convective boundary, such that the molecular opacity dominates and the scaling can be approximated with our default parameters listed in Table \ref{table:default_parameters}. The most important argument for the lack of grains is that mass deposition by both pebbles and planetesimals is expected to occur predominantly in the deep interior, below the radiative-convective boundary \citepalias{Brouwers2018}. Smaller grains that enter with the disk gas can efficiently coagulate and settle \citep{Ormel2014, Mordasini2014}. A caveat to this scenario is that we also neglect the opacity contribution associated with the pebbles themselves as they sediment, a factor that can become especially significant in the outer disk.

Most of the luminosity prior to runaway gas accretion is generated by the conversion of gravitational energy from impacting objects \citep[e.g.,][]{Pollack1996,Hubickyj2005}:
\begin{equation}\label{eq:L}
    L = \frac{G M_\mathrm{c} \dot{M_\mathrm{z}}}{r_\mathrm{c}},
\end{equation}
where $ M_\mathrm{c}$, $r_\mathrm{c}$ are the core mass and radius. Together, Eqs. \ref{eq:nabla_rad}, \ref{eq:opacity_law} and \ref{eq:L} determine the density at the radiative-convective boundary during accretion:
\begin{subequations}
\begin{align}
    \rho_\mathrm{rcb} &= \frac{64\pi\sigma T_\mathrm{disk}^4 r'_\mathrm{B}}{3 \kappa_\mathrm{rcb} L} \label{eq:rho_rcb_pre}\\
    &= {\left(\frac{64\pi\sigma\mu_\mathrm{xy}(\gamma_\mathrm{xy}-1) M_\mathrm{p} r_\mathrm{c}}{3 \kappa_\mathrm{cst} \dot{M}_\mathrm{z} k_\mathrm{B} \gamma_\mathrm{xy} M_\mathrm{c}}\right)}^\frac{1}{1+\beta} T_\mathrm{disk}^\frac{3-\delta}{1+\beta}, \label{eq:rho_rcb}
\end{align}
\end{subequations}
where $r'_\mathrm{B} = \frac{\gamma_\mathrm{xy}-1}{\gamma_\mathrm{xy}} r_\mathrm{B}$ is a modified Bondi radius. If we substitute our default values from Table \ref{table:default_parameters} for a grain-free envelope, we find that $\rho_\mathrm{rcb, gf} \gg \rho_\mathrm{disk}$, unless the planet is both exceedingly small and located close-in, meaning that the radiative zone is important. If a larger, ISM-like opacity is substituted, the envelope reduces to two convective zones, with the outer boundary $r_\mathrm{rcb} = r_\mathrm{out}, \rho_\mathrm{rcb} = \rho_\mathrm{disk}$.

\subsection{Intermediate convective region}\label{sect:intermediate region}
Interior to the radiative zone sits a metal-free intermediate convective region. We assume an adiabatic gradient in order to describe heat transport here, so that the pressure and density are related as $ P_\mathrm{g}  \rho_\mathrm{g}^{-\gamma_\mathrm{xy}} = K$. Integrating Eq. \ref{eq:dP_dm} from the radiative boundary conditions defined by Eq. \ref{eq:rho_rcb} and using the ideal gas law yields the physical structure:
\begin{subequations}
\begin{align}
    T(r) &= T_\mathrm{disk} \left(1 + r'_\mathrm{B} \big( \frac{1}{r} - \frac{1}{r_\mathrm{rcb}} \big) \right), \label{eq:T_ad_xy} \\
    \rho(r) &= \rho_\mathrm{rcb} {\left(1 + r'_\mathrm{B} \big( \frac{1}{r} - \frac{1}{r_\mathrm{rcb}} \big) \right)}^\frac{1}{\gamma_\mathrm{xy} - 1}, \label{eq:rho_ad_xy} \\
    P(r) &= P_\mathrm{rcb} {\left(1 + r'_\mathrm{B} \big( \frac{1}{r} - \frac{1}{r_\mathrm{rcb}} \big) \right)}^\frac{\gamma_\mathrm{xy}}{\gamma_\mathrm{xy} - 1}. \label{eq:P_ad_xy}
\end{align}
\end{subequations}
These expressions are identical to those derived by \citet{Ginzburg2016} and differ by a constant factor near unity compared to those by \citet{Piso2014}, who include non-adiabatic effects. The envelope begins to significantly heat up inside this convective zone and the pressure and density curves are consequently less steep than their radiative counterparts. The metal-free intermediate region of our model ends either at the core, or at the outer boundary of the high-Z region, which we discuss in the next subsection.

\subsection{High-Z convective inner region}\label{sect:high-Z region}
The thermal ablation of solids and the absorption of the resultant vapor are temperature-dependent processes that only occur where the envelope is sufficiently hot. The maximum gas mass fraction $Z$ that can locally be occupied by heavy vapor molecules is given by
\begin{equation}\label{eq:vapor_fraction}
Z = \frac{\mu_{Z}}{\mu_\textrm{g}} \frac{P_\mathrm{sat}}{P_\mathrm{g}}
,\end{equation}
where $P_\mathrm{sat}, P_\mathrm{g}$ are the saturation vapor and hydrostatic pressures and $\mu_\mathrm{z}, \mu_\mathrm{g}$ are the high-Z and local mean molecular weights. The latter depends on the composition and can be written as
\begin{subequations}
\begin{align}
    \frac{1}{\mu_\mathrm{g}} &= \frac{1-Z}{\mu_\mathrm{xy}} + \frac{Z}{\mu_\mathrm{z}}, \label{eq:mu_mix}\\
    \mu_\mathrm{g} &\simeq \frac{\mu_\mathrm{xy}}{1-Z} \; \; \mathrm{if} \; \;  Z < \frac{\mu_\mathrm{z}}{\mu_\mathrm{z}+\mu_\mathrm{xy}}. \label{eq:mu_mix_2}
\end{align}
\end{subequations}
The exponential nature of the vapor pressure \citep[e.g.,][]{Stull1947, Melosh2007, Kraus2012} means that the compositional gradient indicated by Eq. \ref{eq:vapor_fraction} is very steep once the local temperature exceeds the required value to absorb evaporated solids. We use this rationalization to simplify the transition from a metal-free ulterior to a vapor-containing interior with a step-function at radius $r_\mathrm{vap}$, determined by the depth where $T_\mathrm{g} = T_\mathrm{vap}$. The physical value of $T_\mathrm{vap}$ depends on the composition of pollutants, as well as on the envelope mass that determines the hydrostatic pressure in Eq. \ref{eq:vapor_fraction}. In our model, it is essentially a scaling variable, and we set its default value equal to 2500 K, an estimate for silicate vapor. We use a differently scaled value in Sect. \ref{sect:Rainout} to show that our model can match previously obtained results by \citetalias{Brouwers2018} if we fit this parameter.

Simulations by \citet{Bodenheimer2018} have shown that a second consequence of the compositional gradient is a steep increase in the temperature at the boundary of the high-Z region, as according to the Ledoux criterion convection can locally be prevented by its stabilizing influence. We cannot quantitatively account for this effect in our analytical model, and instead assume the temperature to be continuous between the two convective regions. This is a significant simplification that is also present in \citetalias{Brouwers2018}, although it is likely second order to the inaccuracies introduced by the use of ideal gases.

The simplified compositional jump of our model means that we can write the density ($\rho_\mathrm{vap}$) and location ($r_\mathrm{vap}$) of the high-Z region's outer boundary as
\begin{subequations}
\begin{align}
    \rho_\mathrm{vap} &= {\left(\frac{T_\mathrm{vap}}{T_\mathrm{disk}}\right)}^\frac{1}{\gamma_\mathrm{xy} - 1} \frac{\mu_\mathrm{g}}{\mu_\mathrm{xy}} \rho_\mathrm{rcb}, \label{eq:rho_step} \\
    r_\mathrm{vap} &= \frac{r'_\mathrm{B}}{(T_\mathrm{vap}/T_\mathrm{disk} - 1) + r'_\mathrm{B}/r_\mathrm{rcb} }. \label{eq:r_step}
\end{align}
\end{subequations}
The pressure remains continuous, as the envelope stays in hydrostatic equilibrium. Analogous to our derivation of Eqs. \ref{eq:T_ad_xy} - \ref{eq:P_ad_xy}, we use this pressure continuity to find its thermodynamic properties. We integrate Eq. \ref{eq:dP_dm} with the boundary conditions of Eqs. \ref{eq:rho_step} - \ref{eq:r_step}, where the pressure and temperature are still linked adiabatically, but with a lower entropy. This yields the conditions interior to $r_\mathrm{vap}$:
\begin{subequations}
\begin{align}
    T(r) &= T_\mathrm{vap} \left(1 + r''_\mathrm{B} \big( \frac{1}{r} - \frac{1}{r_\mathrm{vap}} \big) \right), \label{eq:T_ad_z}\\
    \rho(r) &= \rho_\mathrm{vap} {\left(1 + r''_\mathrm{B} \big( \frac{1}{r} - \frac{1}{r_\mathrm{vap}} \big) \right)}^\frac{1}{\gamma_\mathrm{g} - 1}, \label{eq:rho_ad_z}\\
    P(r) &= P_\mathrm{sat} {\left(1 + r''_\mathrm{B} \big( \frac{1}{r} - \frac{1}{r_\mathrm{vap}} \big) \right)}^\frac{\gamma_\mathrm{g}}{\gamma_\mathrm{g} - 1}, \label{eq:P_ad_z}
\end{align}
\end{subequations}
where $\gamma_\mathrm{g}$ is the adiabatic index of the high-Z region and $r''_\mathrm{B} = \frac{\gamma_\mathrm{g}-1}{\gamma_\mathrm{g}} \big(\mu_\mathrm{g}/\mu_\mathrm{xy}\big) \big(T_\mathrm{disk}/T_\mathrm{vap}\big) r_\mathrm{B}$ is a second modified Bondi radius, similar to $r'_\mathrm{B}$ in the metal-free region. 

\subsubsection{Model simplifications}
Evaluating the value of $\gamma_\mathrm{g}$ in the high-Z region is a complex task and it can vary widely depending on the composition and thermodynamic conditions. If one just considers the degrees of freedom of the constituent molecules, a pure silicate (SiO$_2$) vapor interior is characterized by $\gamma \sim 1.2$, while a neutral hydrogen gas lies closer to $\gamma = 1.4$. However, as shown by \citet{Bodenheimer2018} and \citetalias{Brouwers2018}, the interior of a polluted planet becomes characterized by temperatures that exceed $10^4$ K, meaning that hydrogen not only dissociates but partially ionizes, reducing the adiabatic index significantly. Besides, silicate molecules can both dissociate and chemically react at this temperature, making the situation even more complex. No current interior model takes all these processes into account, and the value of $\gamma_\mathrm{g}$ at the interior is therefore deeply uncertain and likely highly variable through a planet's evolution. In order to continue to work with analytical solutions, we have to assume it to be a constant. To reflect the likelihood of the value being less than 4/3, we have chosen characteristic default values of 1.2 and 1.25 for pure silicate and mixed compositions respectively. But we note that the constancy of $\gamma$ is a key assumption of the model.

We further note that while the Bondi radius scales with the total mass in our model, the structure Eqs. \ref{eq:T_ad_z} - \ref{eq:P_ad_z} do not fully self-consistently account for the self-gravity of the envelope. This is a fundamental limitation that characterizes all analytical models, and has been shown by \citet{Piso2014} to be a significant effect, even when the envelope mass is an order of magnitude smaller than that of the core. Since we consider planetary evolution up to the critical mass, this provides another limitation to our quantitative predictions. However, compared to \citet{Piso2014}, the consequences are less severe, because in our model the polluted interior is characterized by $\gamma_\mathrm{g} < \frac{4}{3}$ compared to $\gamma_\mathrm{g} = 1.4$ in the model by \citet{Piso2014}. This difference means that most of the envelope mass in our model is located close to the core surface, instead of near the outer boundary. Consequently, most of the envelope feels a gravitational pull that is approximately constant as a function of radius.

\begin{figure}[t!] 
\centering
\includegraphics[width=\hsize]{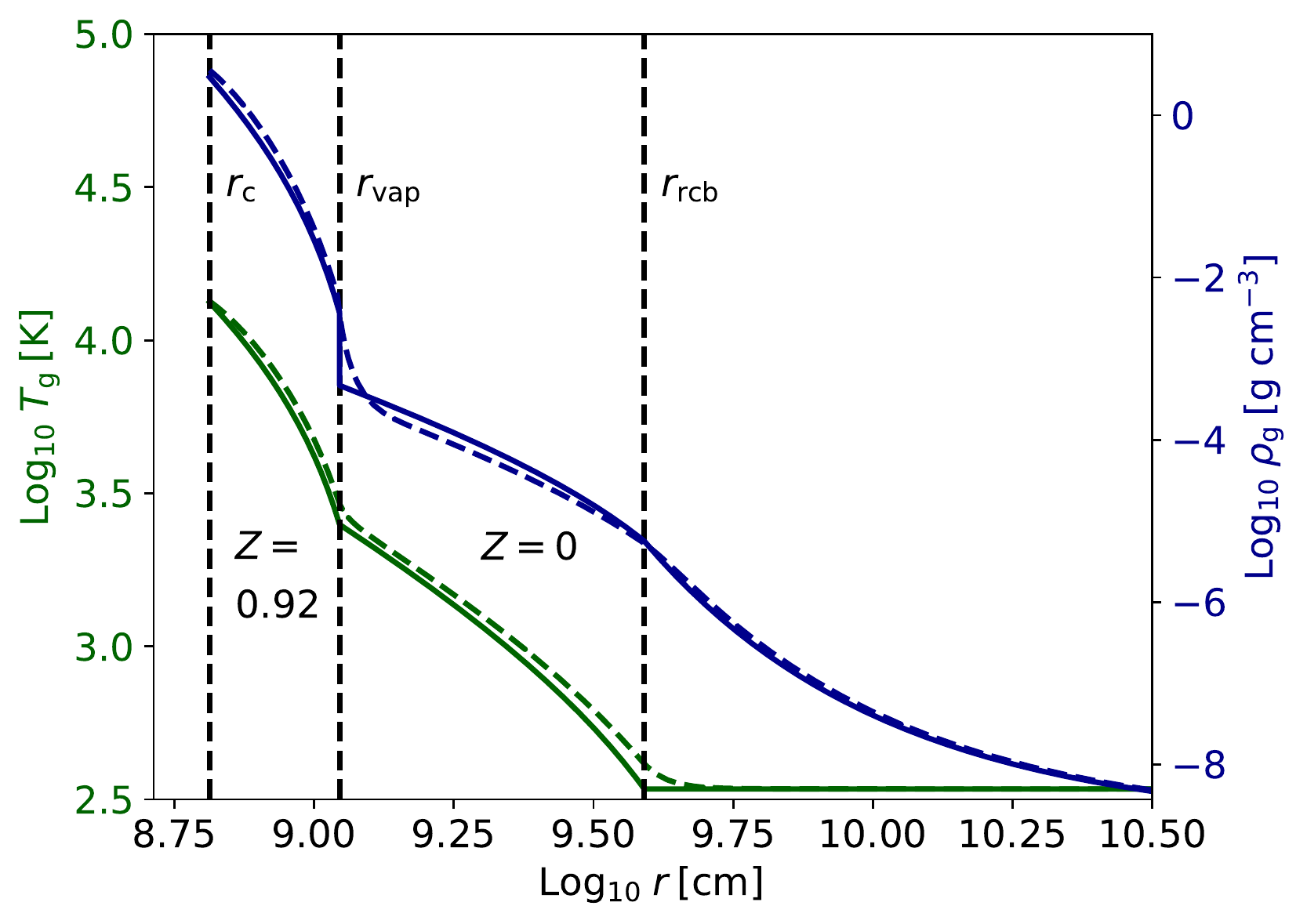} 
\caption{Envelope profiles of temperature (green) and density (blue) curves of an Earth-mass planet with $M_\mathrm{c} = 0.62 \; \mathrm{M_\oplus}$ at 1 AU. The dashed curves are obtained numerically from the structure equations, whereas the solid curves indicate our analytical model, plotted with the default parameters of Table 1. In our analytical model, the entropy and density change as a step-function at $r_\mathrm{vap}$, while the temperature and pressure remain continuous. \label{fig:interior_conditions}}
\end{figure}

To illustrate the interior structure of our analytical model, Fig. \ref{fig:interior_conditions} shows the temperature and density for an Earth-mass forming planet with  $M_\mathrm{c} = 0.62 \; \mathrm{M_\oplus}$ at 1 AU, using the default parameters from Table \ref{table:default_parameters}. Our analytical model iteratively calculates the vapor mass fraction to be $Z=0.92$ in the envelope's interior. The figure also shows comparison to a simulation with the same parameters where the structure equations are numerically integrated with scientific python’s
(scipy) bvp\textunderscore solve routine. The figure shows that the assumption of an isothermal outer region is well justified for a grain-free envelope. The steep compositional gradient is simplified to a compositional jump at $r_\mathrm{vap}$ that captures the rapidly increasing density.

\subsubsection{The assumption of uniform mixing}\label{sect:uniform_mixing}
Physically, the most important model assumption we make is that the convective regions are perfectly mixed, meaning that the compositional mass fractions are uniform in each zone. This significantly simplifies the interior conditions described by Eqs. \ref{eq:T_ad_z} - \ref{eq:P_ad_z}, and allows us to tackle the problem analytically. The assumption of perfect mixing makes our interior model somewhat comparable to the numerical works by \citet{Venturini2015, Venturini2016}, and contrasts with the work of \citet{Bodenheimer2018}, who consider the opposite case without compositional mixing and always find a saturated interior. Unfortunately, it is currently unclear how efficiently mixing proceeds in forming planets, an issue that stems from the poorly constrained diffusive constant in planetary interiors \citep{Hori_Ikoma_2011}.

\begin{figure}[t!] 
\centering
\includegraphics[width=\hsize]{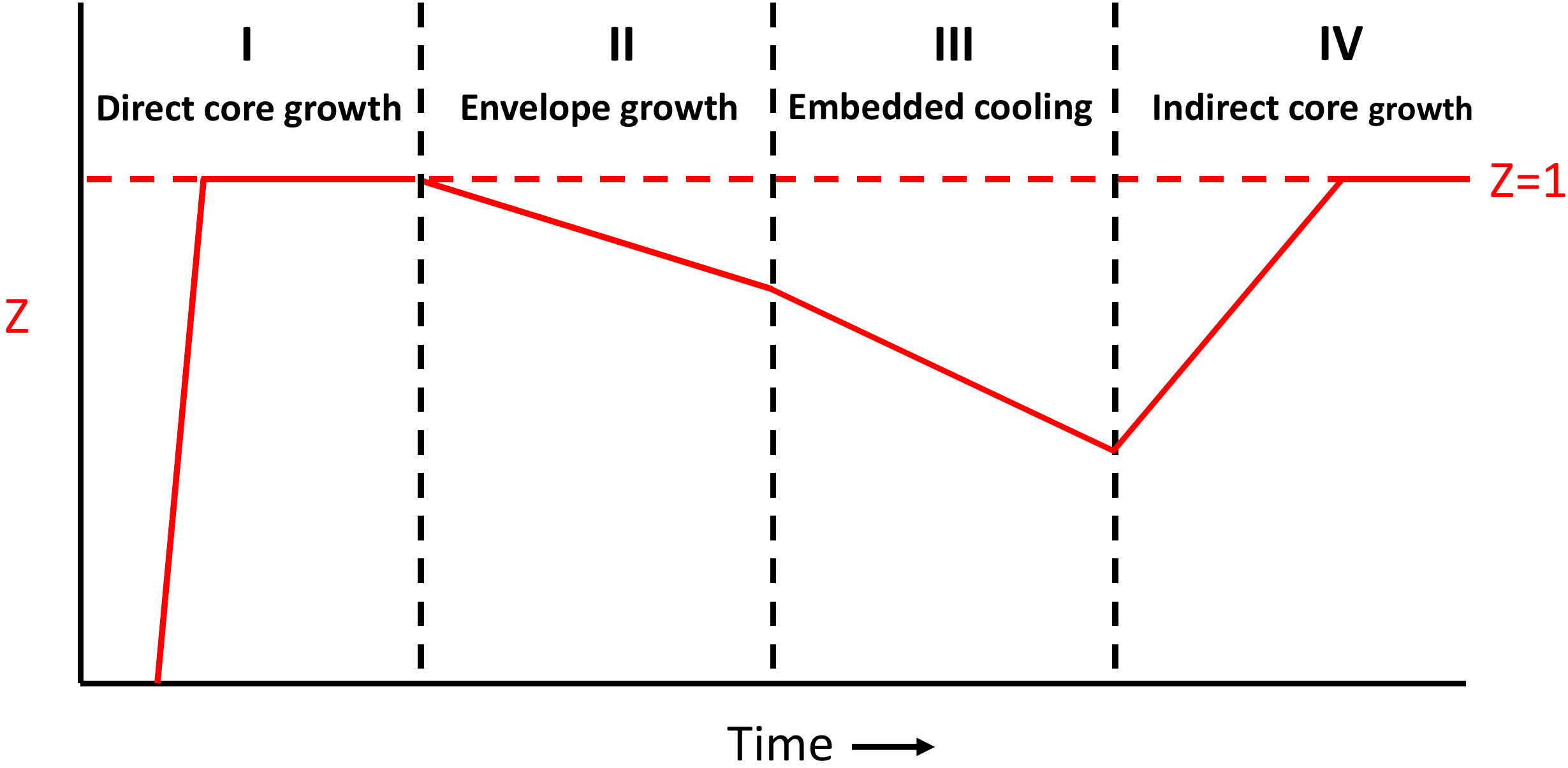} 
\caption{Sketch of the inner envelope's saturation level ($Z$) in the envelope's deep interior during the four evolutionary stages. The planet is initially saturated (phase I), but dilutes as nebular gas flows in (phases II and III). The envelope can eventually become saturated again if it cools down sufficiently (phase IV). \label{fig:sketch_Z}}
\end{figure}
We sketch in Fig. \ref{fig:sketch_Z} how the vapor mass fraction in our model varies during the different evolutionary stages. Initially, the interior is fully saturated with vapor ($Z=1$, phase I). This changes when the planet becomes so massive that the interior can absorb more vapor than is available from the accretion of solids (II). The interior dilutes further still if the planet subsequently goes through a phase of embedded cooling (III). Eventually though, the envelope can return to saturation if the interior cools down sufficiently in isolation (IV). The next sections are devoted to analyzing and working out these phases in more detail.

\section{Phase I: Direct core growth}\label{sect:phase_I}
Young forming planets that have yet to accrete significant mass are only able to bind tenuous envelopes with temperatures similar to that of the surrounding disk. As a result, impactors initially have no difficulty passing through and reaching the core. But as a planet gains mass, its interior simultaneously heats up and eventually reaches temperatures sufficient to vaporize and absorb some of the incoming solids. While the interior is saturated, the core still grows through rainout of excess material, a process that slows down as the planet grows and eventually halts completely \citepalias{Brouwers2018}. In this section, we first derive an analytical expression for the minimal sizes that accreting pebbles require to reach the core. We then consider the criterion for complete absorption, indicating the end of direct core growth, labeled as phase I in our model. Finally, we show that we can reproduce the numerically obtained core masses from \citetalias{Brouwers2018} if we tune the parameter $T_\mathrm{vap}$.

\subsection{Impacts}\label{sect:impacts}
Any impactor that travels through an envelope encounters increased gas densities and consequently experiences friction and drag. For planetesimals that are sufficiently large to maintain high velocities, this can lead to significant ablation and cause compressive fractures (\citealt{Mordasini2015}; \citetalias{Brouwers2018}; \citealt{Valletta2018}). Smaller objects like pebbles are slowed down more efficiently and instead predominantly lose mass through evaporation by thermal radiation in the envelope's deep interior. 

To evaluate a pebble's mass-loss, we first note that because enrichment occurs as a consequence of evaporation, we can model its onset in a metal-free envelope of disk composition. We can also use the quadratic temperature-dependence of radiative heat transport to infer that most of the evaporation will happen close to the core, where the envelope is hottest. Since an impactor's outer temperature is limited to $T_\mathrm{evap}$, the net absorbed energy flux $F_\mathrm{in}$ during the dominant final stage of an impact can be approximated as
\begin{subequations}
\begin{align} \label{eq:P_in}
    F_\mathrm{in} &= 4 \sigma A_\mathrm{i} \big(T_\mathrm{g}^4 - T_\mathrm{evap}^4\big) \\
    &\simeq 4 \sigma A_\mathrm{i} T_\mathrm{g}^4,
\end{align}
\end{subequations}
where $A_\mathrm{i} = \pi {R_\mathrm{i}}^2$ is the impactor's frontal surface area. This simplification is reasonable as long as the pebbles are not too small to instantly vaporize when $T_\mathrm{g} \approx T_\mathrm{evap}$. The local gas temperature and density deep down ($r \ll r_\mathrm{rcb}$) can also be simplified to
\begin{equation}
    T(r) \simeq T_\mathrm{disk} \frac{r'_\mathrm{B}}{r}  \quad , \quad \rho(r) \simeq \rho_\mathrm{rcb} {\left(\frac{r'_\mathrm{B}}{r} \right)}^\frac{1}{\gamma_\mathrm{xy} - 1}. \label{eq:T_rho_ad_xy_approx}
\end{equation}

Provided that the impactors are sufficiently small, they will sediment at speeds $v_\mathrm{sed}$ close to their terminal velocity $v_\mathrm{terminal}$. Pebble-sized impactors in this region are typically characterized by Stokes drag due to the small mean free path of the dense gas \citep{Whipple1972, Weidenschilling1977, Armitage2007}. Equating the gravitational and drag forces yields the terminal velocity
\begin{equation}\label{eq:v_eq}
    v_\mathrm{sed} \approx v_\mathrm{terminal} = {\left(\frac{2 G M_\mathrm{c}}{C_\mathrm{d} \rho_\mathrm{g} r^2} \frac{M_\mathrm{i}}{A_\mathrm{i}}\right)}^\frac{1}{2},
\end{equation}
where we assume spherical impactors with mass $M_\mathrm{i}$ and density $\rho_\mathrm{i}$. Combining Eqs. \ref{eq:P_in} - \ref{eq:v_eq} and neglecting the outer integration boundary because most evaporation happens close to the core, we find that an impactor absorbs an amount of energy $E_\mathrm{in}$ during its trajectory, approximately equal to
\begin{subequations}
\begin{align}
    E_\mathrm{in} &= -\int_{r_\mathrm{out}}^{r_\mathrm{c}} \frac{F_\mathrm{in}}{v_\mathrm{sed}} dr \label{eq:E_in_int}\\
    &\simeq \frac{2(\gamma_\mathrm{xy}-1)}{4\gamma_\mathrm{xy}-3} \sigma T^4_\mathrm{disk}  {\left( \frac{8 C_\mathrm{d} \rho_\mathrm{rcb} {A_\mathrm{i}}^3}{G M_\mathrm{c} M_\mathrm{i}} \right)}^\frac{1}{2} {r'_\mathrm{B}}^\frac{8\gamma_\mathrm{xy}-7}{2(\gamma_\mathrm{xy}-1)}  r_\mathrm{c}^\frac{3-4\gamma_\mathrm{xy}}{2(\gamma_\mathrm{xy}-1)}. \label{eq:E_in}
\end{align}
\end{subequations}
Impactors can only fully ablate if they absorb sufficient energy to vaporize all their material, set by $E_\mathrm{req} = M_\mathrm{i} u_\mathrm{evap}$. Per definition, we can say that this happens when their radius is smaller than some critical radius ($R_\mathrm{i} < R_\mathrm{crit}$), which we can now estimate from Eq. \ref{eq:E_in} to be
\begin{equation}\label{eq:R_crit}
\begin{split}
    R_\mathrm{crit} &\simeq  \frac{1}{\rho_\mathrm{i}} {\left(\frac{8C_\mathrm{d}\rho_\mathrm{rcb}}{G}\right)}^\frac{1}{3} {\left(\frac{3\sigma(\gamma_\mathrm{xy}-1)}{2u_\mathrm{evap}(4\gamma_\mathrm{xy}-3)}\right)}^\frac{2}{3} {\left(\frac{3}{4 \pi \rho_\mathrm{c}} \right)}^\frac{3-4\gamma_\mathrm{xy}}{9(\gamma_\mathrm{xy}-1)} \\
    & {\left(\frac{(\gamma_\mathrm{xy}-1)G\mu_\mathrm{xy}}{\gamma_\mathrm{xy} k_\mathrm{B}} \right)}^\frac{8\gamma_\mathrm{xy}-7}{3(\gamma_\mathrm{xy}-1)}
    {T_\mathrm{disk}}^\frac{-1}{3(\gamma_\mathrm{xy}-1)} {M_\mathrm{c}}^\frac{17\gamma_\mathrm{xy}-15}{9(\gamma_\mathrm{xy}-1)}. \\
\end{split}
\end{equation}
Equation \ref{eq:R_crit} can be simplified further by substituting our default parameters from Table \ref{table:default_parameters}:
\begin{equation}\label{eq:R_crit_simple}
    R_\mathrm{crit} \approx 596 \: \mathrm{cm} \: {\left(\frac{d}{\mathrm{AU}}\right)}^\frac{10}{27} {\left(\frac{M_\mathrm{c}}{\mathrm{M}_\mathrm{\oplus}}\right)}^{2.45}.
\end{equation}
This result shows that impactors must be larger than m-sized in order to reach the cores of forming Earth-mass planets, regardless of the planet's location in the disk. It indicates the necessity of accounting for the ablated material in pebble accretion and is consistent with the numerical results of impact simulations by \citetalias{Brouwers2018} and with our new, simpler impact model presented in Appendix B.

\subsection{Rainout}\label{sect:Rainout}
\begin{figure}[t!] 
\centering
\includegraphics[width=\hsize]{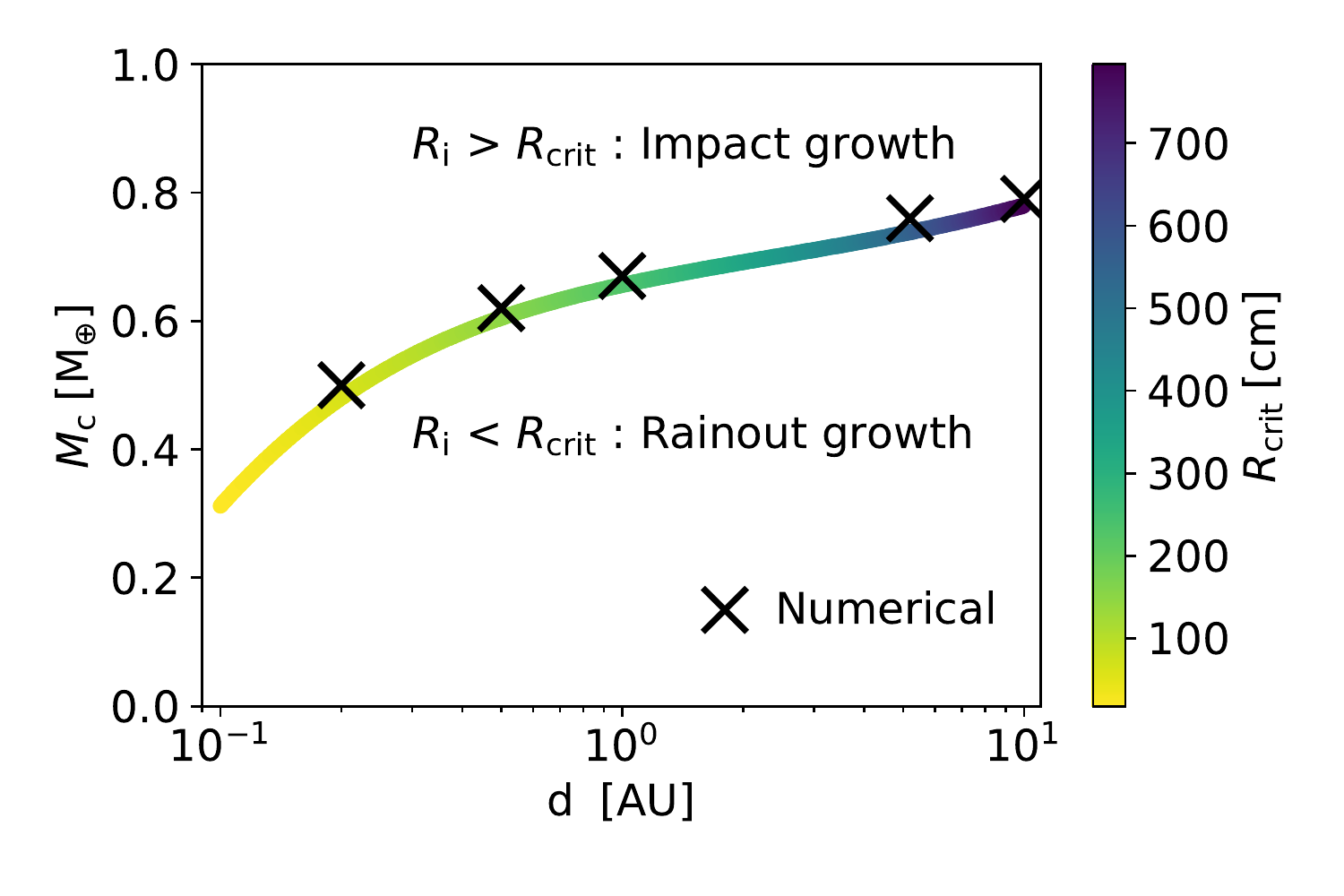} 
\caption{Final core masses of direct core growth (phase I). The line indicates our semi-analytical calculation described in the text. The numerical simulations of \citetalias{Brouwers2018} are indicated by the black crosses.\protect\footnotemark Impacts and rainout determine the end of direct core growth above and below the line, respectively. Its color indicates the impactor size necessary to grow the core beyond this value. \label{fig:core_growth}} 
\end{figure}
\footnotetext{The simulations in \citetalias{Brouwers2018} used an opacity that was a factor 10 larger than those from \citet{Bell1994} due to a conversion error. The values in the graph correspond to runs with corrected opacity.}
As a planet accretes mass, its hot interior region grows and expands outwards, continually being filled with vapor from evaporated impactors (\citealt{Iaroslavitz2007}; \citetalias{Brouwers2018}; \citealt{Bodenheimer2018}). We model this in the same way as \citetalias{Brouwers2018}, where the fraction that can be accommodated as vapor is initially limited and the excess is forced to rain out. We begin by estimating the mass of this vapor region. Because the adiabatic index of the interior region is less than $\frac{4}{3}$ in our model, its mass is mainly located at the core boundary and can initially be approximated as
\begin{subequations}
\begin{align}
    M_\mathrm{z, env} &= 4 \pi \int^{r_\mathrm{vap}}_{r_\mathrm{c}} {\rho_\mathrm{g} r^2 dr} \\
    &\approx \pi \rho_\mathrm{cg} r_\mathrm{c}^2 \big( r_\mathrm{vap} - r_\mathrm{c} \big), \label{eq:M_env_Z_approx}
\end{align}
\end{subequations}
where $\rho_\mathrm{cg}$ is the density of the gas at the core-envelope boundary and follows from Eq. \ref{eq:rho_ad_z}. The expansion of the high-Z region accelerates until it reaches the point that $\dot{M}_\mathrm{z, env} = \dot{M}_\mathrm{z}$ and the complete solids flux is absorbed. This is equivalent to the condition that the core stops growing. The low-mass regime makes it difficult to derive a comprehensive analytical expression corresponding to the end of direct core growth, but we can semi-analytically determine it from Eq. \ref{eq:M_env_Z_approx} by locating the mass where the condition $\dot{M}_\mathrm{z, env} = \dot{M}_\mathrm{z}$ is reached.

In order to show that we can reproduce the results of \citetalias{Brouwers2018} with our model, we use the same empirical expression of $P_\mathrm{sat}$ from \citet{Stull1947} for general silicate rock. We set the outer boundary of the high-Z radius at the depth where the vapor pressure first approaches the total pressure, such that $Z = 0.1$ in Eq. \ref{eq:vapor_fraction}. The resulting core mass is plotted in Fig. \ref{fig:core_growth}, along with a comparison to the values we found using purely numerical methods in \citetalias{Brouwers2018}. 

When we substitute the resultant core masses into Eq. \ref{eq:R_crit_simple}, we see that impactors can still pass through these envelopes if they are larger than m-sized. Note that these results only apply to silicate pebbles, and that in the outer disk it is instead predominantly ice that will sublimate. Its evaporation occurs at a lower temperature, making the picture more complex if impactors of different compositions are taken into account. To account for the possibility of a planet accreting larger objects like planetesimals, and because small cores combined with massive ideal-gas polluted envelopes can lead to nonphysical densities, we choose to parameterize the core mass in the subsequent sections of this work. An analogous treatment was conducted by Venturini (2016) to account for unknown properties of accreted planetesimals.

\section{Phase II: Envelope growth}\label{sect:phase_II}
Any material that a planet accretes beyond its initial phase of direct core growth becomes part of its envelope. But due to the rapidly growing hot interior, this stream eventually becomes insufficient to keep the envelope saturated and metal-poor nebular gas begins to flow in. The vapor mass fraction of the high-Z region consequently drops below unity and the thermodynamic gradients flatten. We refer to this period of the planet's formation as phase II.

\subsection{Envelope mass comparison}
For the sake of comparison, we first consider the classical case of a metal-free envelope. Their convective part typically dominates over the tenuous radiative zone in mass ($M_\mathrm{xy} \approx M_\mathrm{xy, intermediate}$). After substituting the density relation of Eq. \ref{eq:rho_ad_xy}, it can be written as
\begin{subequations}
\begin{align}
    M_\mathrm{xy,intermediate} &= 4\pi \int_{r_\mathrm{c}}^{r_\mathrm{rcb}} {\rho(r) r^2 dr} \\
    &\simeq
    4\pi\frac{\gamma_\mathrm{xy}-1}{3\mathrm{\gamma_\mathrm{xy}}-4} \rho_\mathrm{rcb} {r'}_\mathrm{B}^\frac{1}{\gamma_\mathrm{xy}-1} r_\mathrm{rcb}^\frac{3\gamma_\mathrm{xy}-4}{\gamma_\mathrm{xy}-1}, \label{eq:M_env_XY}
\end{align}
\end{subequations}
where we twice used that $r_\mathrm{rcb} \gg r_\mathrm{c}$, before and after the integration. Since $\gamma_\mathrm{xy}>\frac{4}{3}$, the majority of the envelope mass is located near the radiative-convective boundary. This means that Eq. \ref{eq:M_env_XY} is suitable to approximate both the total envelope mass of a metal-free envelope, as well as the mass of the intermediate region in the case of a polluted envelope.

The difference, as explained previously, is that these polluted envelopes have an additional layer of high-Z vapor that surrounds their cores. Similar to the equation above, we integrate the density profile and find that it can be written as
\begin{subequations}
\begin{align}
    M_\mathrm{env, inner} &= 4\pi \int_{r_\mathrm{c}}^{r_\mathrm{vap}} {\rho_\mathrm{g}(r) r^2 dr} \label{eq:M_env_int}\\
    &\simeq 4\pi\frac{\gamma_\mathrm{g}-1}{4-3\gamma_\mathrm{g}} \rho_\mathrm{vap} {r''}_\mathrm{B}^\frac{1}{\gamma_\mathrm{g}-1} r_\mathrm{c}^\frac{3\gamma_\mathrm{g}-4}{\gamma_\mathrm{g}-1}, \label{eq:M_env_approx}
\end{align}
\end{subequations}
where we evaluated Eq. \ref{eq:M_env_approx} in the limit $r_\mathrm{vap} \gg r_\mathrm{c}$, a reasonable approximation when the envelope contains a significant fraction of the metals as vapor (since $\rho_\mathrm{g} < \rho_\mathrm{c}$). Because $\gamma_\mathrm{g} <\frac{4}{3}$ in our model, most of the vapor is concentrated around the core. We plot the hydrogen-helium component of both the polluted and metal-free envelopes as a function of the metal mass in Fig. \ref{fig:mass_metal-poor}. The figure shows that the accretion of nebular gas rapidly accelerates when the envelope begins to absorb the silicates as vapor, even if the core contains the majority of the metals. This means that the total mass of polluted envelopes can typically be approximated by just that of the inner high-Z region. Second, it shows that well-mixed polluted envelopes reach runaway accretion sooner than their metal-free counterparts, as they are able to suck in more nebular gas with the same amounts of solids accreted. Physically, this is because the presence of vapor actively raises the gas density by lowering the adiabatic index and increasing the mean molecular weight \citep{Stevenson1982, Wuchterl1993, Venturini2015, Venturini2016}.

\subsection{Critical mass determination}\label{sect:runaway_analytical}
\begin{figure}[t!] 
\centering
\includegraphics[width=\hsize]{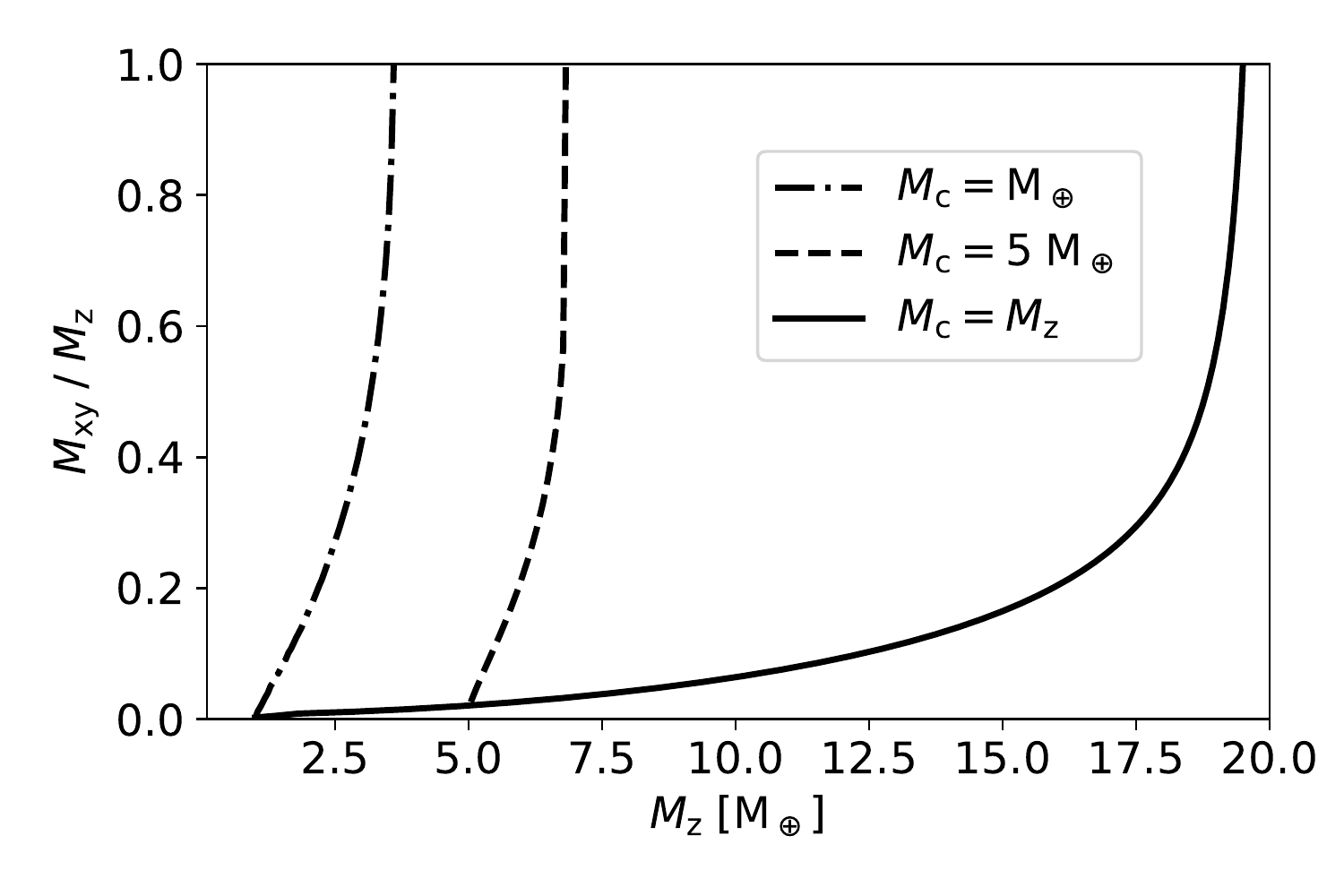}
\caption{Evolution of three planets in phase II with our default parameters (see Table \ref{table:default_parameters}) at 0.1 AU. The solid curve indicates the classical case, where the core grows indefinitely and the envelope is metal-free. In the dash-dotted and dashed lines, the growth of the core is limited to $\mathrm{M_\oplus}$ and 5 $\mathrm{M_\oplus}$, respectively. Accretion of nebular gas accelerates when the envelope becomes polluted. This leads to a smaller mass in metals at the onset of runaway accretion.}
\label{fig:mass_metal-poor}
\end{figure}
Next, we derive an approximate criterion for the onset of runaway gas accretion for polluted planets. Instead of turning to the common method of looking for a maximum in the core mass, which is not directly coupled to the total mass in our model and is thus not possible, we simply approximate runaway accretion to initiate around the crossover mass when $M_\mathrm{z} = M_\mathrm{xy}$. Figure \ref{fig:mass_metal-poor} shows this to be a good approximation, as the envelope mass increases steeply as a function of the metal mass near this point.

Our first step is to parameterize the pollution fraction $f_\mathrm{z} = \big(M_\mathrm{z} - M_\mathrm{c}\big)/M_\mathrm{z}$, to allow the core, envelope and total mass at runaway to be expressed as a function of the metal mass:
\begin{equation}
    M_\mathrm{c} = (1-f_\mathrm{z}) M_\mathrm{z} \; , \;
    M_\mathrm{env} = (1+f_\mathrm{z}) M_\mathrm{z} \; , \;
    M_\mathrm{p} = 2 M_\mathrm{z}.
\end{equation}
Similarly, the mean molecular weight can also be written in terms of the pollution fraction as (see Eq. \ref{eq:mu_mix_2})
\begin{subequations}
\begin{align}
    \mu_\mathrm{g} &\simeq \frac{\mu_\mathrm{xy}}{1 - Z_\mathrm{env,crit}} \\
    &\simeq (1+f_\mathrm{z}) \mu_\mathrm{xy}. \label{eq:mu_approx}
\end{align}
\end{subequations}
Together, Eqs. \ref{eq:M_env_approx} - \ref{eq:mu_approx} directly yield an analytical approximation of the critical metal mass; the maximum mass in solids that a planet can accrete before it enters runaway accretion:
\begin{equation}
    M_\mathrm{z,crit} = K(\gamma_\mathrm{g}) \rho_\mathrm{rcb}^{-{\frac{3(\gamma_\mathrm{g}-1)}{2}}} T_\mathrm{disk}^\frac{3(\gamma_\mathrm{g}-1)}{2(\gamma_\mathrm{xy}-1)} T_\mathrm{vap}^\frac{3(\gamma_\mathrm{xy}-\gamma_\mathrm{g})}{2(\gamma_\mathrm{xy}-1)} \rho_\mathrm{c}^\frac{3\gamma_\mathrm{g}-4}{2} {(1+f_\mathrm{z})}^{-\frac{3}{2}} (1-f_\mathrm{z})^\frac{4-3\gamma_\mathrm{g}}{2},
\end{equation}
where $K(\gamma_\mathrm{g})$ is a constant pre-factor, given by
\begin{equation}
    K(\gamma_\mathrm{g}) = {\left(4\pi \frac{\gamma_\mathrm{g}-1}{4-3\gamma_\mathrm{g}}\right)}^{-\frac{3(\gamma_\mathrm{g}-1)}{2}} {\left(2\frac{\gamma_\mathrm{g}-1}{\gamma_\mathrm{g}} \frac{G\mu_\mathrm{xy}}{k_\mathrm{B}}\right)}^{-\frac{3}{2}} {\left(\frac{3}{4\pi}\right)}^\frac{4-3\gamma_\mathrm{g}}{2}.
\end{equation}
\begin{figure}[t!] 
\centering
\includegraphics[width=\hsize]{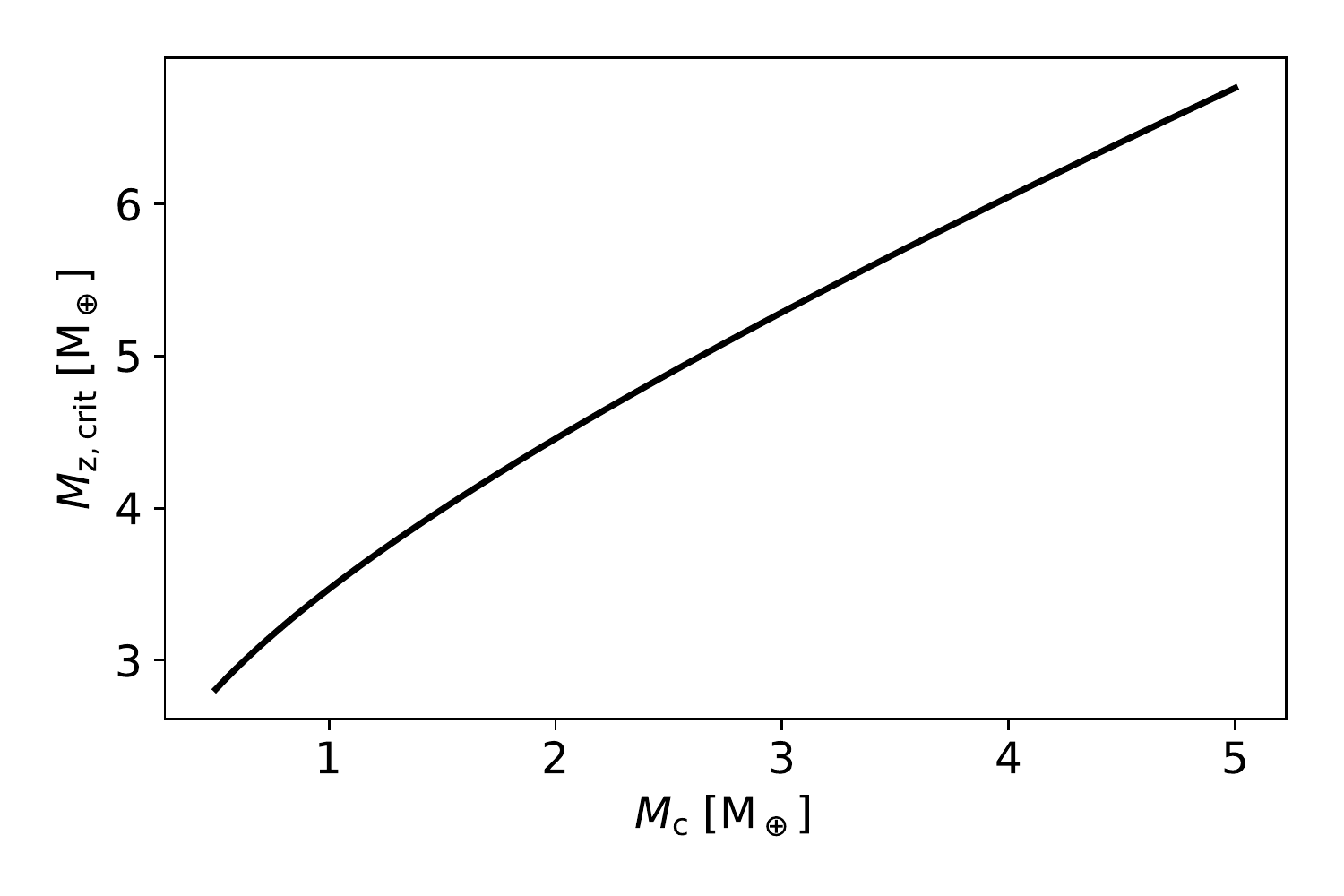}
\caption{The dependence of the critical metal mass on that of the core, with our default parameters (see Table \ref{table:default_parameters}) at 0.1 AU. Smaller cores signify more severely polluted envelopes, and consequently lead to a lower critical metal mass (see Eqs. \ref{eq:M_crit_analytical_2}, \ref{eq:M_crit_analytical_3}).}
\label{fig:critical_z_mass}
\end{figure}
From Sect. \ref{sect:radiative region}, we can substitute $\rho_\mathrm{rcb}$ with Eq. \ref{eq:rho_rcb_pre} and instead introduce $\kappa_\mathrm{rcb}$ as a free parameter:
\begin{align}
    M_\mathrm{z,crit} &= B(\gamma_\mathrm{g}) T_\mathrm{disk}^{\frac{3(4-3\gamma_\mathrm{xy})(\gamma_\mathrm{g}-1)}{(\gamma_\mathrm{xy}-1)(\gamma_\mathrm{g}+1)}} 
    T_\mathrm{vap}^\frac{3(\gamma_\mathrm{xy}-\gamma_\mathrm{g})}{(\gamma_\mathrm{xy}-1)(\gamma_\mathrm{g}+1)} \rho_\mathrm{c}^\frac{4\gamma_\mathrm{g}-5}{\gamma_\mathrm{g}+1}
    {\left(\kappa_\mathrm{rcb} \dot{M}_\mathrm{z}\right)}^\frac{3(\gamma_\mathrm{g}-1)}{\gamma_\mathrm{g}+1}    \\ 
    & \qquad \qquad 
    {(1+f_\mathrm{z})}^{-\frac{3}{\gamma_\mathrm{g}+1}} (1-f_\mathrm{z})^\frac{2-\gamma_\mathrm{g}}{\gamma_\mathrm{g}+1}, \nonumber
\end{align}
where
\begin{align}
    B(\gamma_\mathrm{g}) &= {\left(\frac{512\pi^2\sigma(\gamma_\mathrm{g}-1)(\gamma_\mathrm{xy}-1)}{3(4-3\gamma_\mathrm{g})\gamma_\mathrm{xy}}\right)}^{-\frac{3(\gamma_\mathrm{g}-1)}{\gamma_\mathrm{g}+1}}
    {\left(\frac{3}{4\pi}\right)}^\frac{5-4\gamma_\mathrm{g}}{\gamma_\mathrm{g}+1} \\
    & \qquad \qquad 
    {\left(2G\frac{\gamma_\mathrm{g}-1}{\gamma_\mathrm{g}}\right)}^{-\frac{3}{\gamma_\mathrm{g}+1}}
    {\left(\frac{k_\mathrm{B}}{\mu_\mathrm{xy}}\right)}^\frac{3\gamma_\mathrm{g}}{\gamma_\mathrm{g}+1}.
    \nonumber
\end{align}
This result is very generally applicable, but not incredibly insightful due to the various dependencies. We can simplify it significantly by substituting our default parameters from Table \ref{table:default_parameters}. After doing so, the core density drops out and the critical metal mass reduces to
\begin{align}
    M_\mathrm{z,crit} &= 15.8 \; \mathrm{M_\oplus} \; 
    {\left(\frac{\kappa_\mathrm{rcb}}{0.01 \; \mathrm{g \; cm^{-2}}}\right)}^\frac{1}{3}
    {\left(\frac{d}{\mathrm{AU}}\right)}^\frac{7}{54}
    {\left(\frac{T_\mathrm{vap}}{2500 \; \mathrm{K}}\right)}^\frac{16}{27}    \label{eq:M_crit_analytical_2}
     \\ 
    & \qquad \qquad  \quad
    {\left(\frac{\dot{M}_\mathrm{z}}{10^{-5} \; \mathrm{M_\oplus} \; \mathrm{yr}^{-1}}\right)}^\frac{1}{3}
    {(1+f_\mathrm{z})}^{-\frac{4}{3}} (1-f_\mathrm{z})^\frac{1}{3}. \nonumber
\end{align}
By plotting the dependence on $f_\mathrm{z}$, we observe that it can roughly be approximated as $g(f_\mathrm{z}) \approx 1-f_\mathrm{z}$ to yield a more concise result:
\begin{align}
    M_\mathrm{z,crit} &\approx 4.0 \; \mathrm{M_\oplus} \; 
    {\left(\frac{\kappa_\mathrm{rcb}}{0.01 \; \mathrm{g \; cm^{-2}}}\right)}^\frac{1}{6}
    {\left(\frac{d}{\mathrm{AU}}\right)}^\frac{7}{108}
    {\left(\frac{T_\mathrm{vap}}{2500 \; \mathrm{K}}\right)}^\frac{8}{27}     \label{eq:M_crit_analytical_3}
     \\ 
    & \qquad \qquad \;\;
    {\left(\frac{\dot{M}_\mathrm{z}}{10^{-5} \; \mathrm{M_\oplus} \; \mathrm{yr}^{-1}}\right)}^\frac{1}{6}
    {\left(\frac{M_\mathrm{c}}{\mathrm{M_\oplus}}\right)}^\frac{1}{2}.
     \nonumber
\end{align}
We plot the dependence of the critical metal mass on the size of the core in Fig. \ref{fig:critical_z_mass}, where we substitute our default parameters of Table \ref{table:default_parameters} and iterate Eqs. \ref{eq:rho_rcb_pre} and \ref{eq:M_crit_analytical_2} to obtain the opacity at the radiative-convective boundary. The figure shows that planets with smaller cores reach runaway accretion at lower metal masses.
The same result is also visualized in Fig. \ref{fig:mass_metal-poor} by the asymptotes in the lines that represent polluted envelopes. For example, when the core mass is limited to $\mathrm{M_\oplus}$, we find a critical metal mass of around 3.5 $\mathrm{M_\oplus}$, consistent with Eqs. \ref{eq:M_crit_analytical_2} and \ref{eq:M_crit_analytical_3}.
The derivation of a criterion for the critical metal mass is one of the main results of this work. Our expressions are generally applicable to all planets with singular-species pollution and well-mixed interiors, a consequence of the fact that the mean molecular weight of the high-Z region can be approximated as $\mu_\mathrm{g} \simeq \frac{\mu_\mathrm{xy}}{1-Z}$ for all pollutants with $\mu_\mathrm{z} \gg \mu_\mathrm{xy}$. Needless to say, the quantitative numbers are crude approximates, which are likely to vary if the assumptions that underpin our analytical model are relaxed.

More significant are the derived dependencies of the critical metal mass. While the positive scaling with opacity, distance and accretion rate are shared with classical expressions for the critical core mass \citep[e.g.,][]{Stevenson1982, Wuchterl1993, Piso2014}, the additional dependence on core mass and $T_\mathrm{vap}$ are unique to the critical metal mass and indicate a physical difference between the two. The $T_\mathrm{vap}$-dependence shows that an envelope attracts more disk gas when its high-Z region is more extended, as is the case when the pollutant is very volatile (i.e. water ice). The dependence on core mass indicates the influence of a planet's level of pollution, where a larger core essentially makes the planet more similar to the classical scenario, where all the solids are assumed to sediment.

Physically, the mechanism for the onset of runaway accretion at the critical metal mass is likely similar to that of the critical core mass. In essence, the self-gravity of the envelope accelerates its gas accretion to an unstable level. This process is accelerated by the increased mean molecular weight and lower adiabatic index of polluted envelopes, but is simultaneously also slowed down by the dilution of the interior as metal-poor disk gas flows in \citep{Hori_Ikoma_2011}. When the planet approaches the critical metal mass, however, the mean molecular weight gradient $\frac{d\mu_\mathrm{g}}{dZ}$ has reduced sufficiently that it cannot prevent the inflow of nebular gas. This mechanism was previously highlighted by \citet{Venturini2016} (see their Figs. 2, 3).

\section{Phase III: embedded cooling post solids accretion}\label{sect:phase_III}
Planets that end their evolution as super-Earths or sub-Neptunes do not experience a period of excessive gas accretion. It is possible that they accrete solids throughout the the disk's lifetime but still do not reach the critical mass. The alternative scenario is that their solids accretion is halted prematurely when they clear out their orbits or when a gap opens at a greater orbital separation that locally halts the pebble drift \citep{Morbidelli2012, Lambrechts2014, Morbidelli2015}. In the latter case, the envelope's puffed up interior will cease to be supported by accretion luminosity and undergo contraction at a more rapid pace. If this process occurs while the primordial disk is still present, nebular gas continues to flow in and keeps the Bondi/Hill sphere filled \citep[e.g.,][]{Papaloizou2005, ColemanEtal2017}. We label this as phase III, to contrast with the final phase IV when a planet cools outside the disk confines, which we describe in the next section.

While this period of a planet's evolution is often referred to as a phase of cooling, its internal temperature does not actually decrease while the planet remains embedded. In fact, it continues to rise as nebular gas continues to flow in. Instead, the cooling is manifested in the planet's decreasing luminosity, which alters the envelope's interior structure. In order to study this change, we first write the luminosity (with Eq. \ref{eq:rho_rcb_pre}) in terms of the local conditions at the radiative-convective boundary as is commonly done \citep[e.g.,][]{LeeEtal2014, Ginzburg2016}
\begin{equation}\label{eq:L_rad_diff}
    L = \frac{64\pi\sigma T_\mathrm{disk}^4 r'_\mathrm{B}}{3 \kappa_\mathrm{rcb} \rho_\mathrm{rcb}}.
\end{equation}
Eq. \ref{eq:L_rad_diff} shows that as a planet cools and its luminosity correspondingly decreases, the radiative-convective boundary moves further inward to a region with higher density. After substituting our opacity scaling (Eq. \ref{eq:opacity_law}), we can write this as
\begin{equation}\label{eq:rho_rcb_cooling}
    \rho_\mathrm{rcb} = {\left(\frac{64\pi\sigma T_\mathrm{disk}^{4-\delta} r'_\mathrm{B}}{3 \kappa_\mathrm{cst} L }\right)}^\frac{1}{1+\beta}.
\end{equation}
To describe an envelope's evolution during embedded cooling, we only need to consider the factors that actually change over time. With Eq. \ref{eq:rho_rcb_cooling}, this can be succinctly expressed as a scaling relation:
\begin{equation}\label{eq:rho_rcb_scaling}
    \frac{\rho_\mathrm{rcb}}{\rho_\mathrm{rcb,0}} = {\left(\frac{M_\mathrm{p}}{M_\mathrm{p,0}}\right)}^\frac{1}{1+\beta} {\left(\frac{L}{L_0}\right)}^{-\frac{1}{1+\beta}},
\end{equation}
where the subscript 0 indicates a quantity at the start of embedded cooling. We argue based on continuity that $L_0$ is equal to the accretion luminosity at the end of phase II, even if solids accretion ends full-stop. If instead we were to suppose that $L_0$ instantly drops due to the disappearance of the accretion term, the planet would partially collapse to a denser state with higher $\rho_\mathrm{rcb}$ (see Eq. \ref{eq:rho_rcb_cooling}). Since contraction releases energy, a rapid collapse would overshoot the luminosity to a higher value, and this is inconsistent with the a priori assumption of a non-continuous decrease in L. Thus, the luminosity must in fact be continuous in the transition between accretion and embedded cooling, with the release of internal energy taking over as the dominant energy source.

\subsection{metal-free envelope cooling inside a disk}\label{sect:metal-poor_cooling}
For the sake of comparison, we begin by considering the embedded cooling of metal-free envelopes that consist entirely of hydrogen and helium. We note from Eq. \ref{eq:M_env_XY} that their mass scales as $M_\mathrm{env} \propto \rho_\mathrm{rcb} M_\mathrm{p}^3$ for any $\gamma_\mathrm{xy}>\frac{4}{3}$. Together with of Eq. \ref{eq:rho_rcb_scaling}, this yields the scaling relation between their mass accretion and luminosity:
\begin{equation}
    \frac{L}{L_0} = {\left(\frac{M_\mathrm{env}}{M_\mathrm{env,0}}\right)}^{-(1+\beta)} {\left(\frac{M_\mathrm{p}}{M_\mathrm{p,0}}\right)}^{4+3\beta}. \label{eq:L_scaling_xy}
\end{equation}
Eq. \ref{eq:L_scaling_xy} shows that while the luminosity initially decreases from $L_0$ as the planet contracts, it reaches a minimum when the envelope mass becomes comparable to the total mass and nebular gas accretion accelerates. By differentiating w.r.t. $M_\mathrm{env}$ and using that $M_\mathrm{p} = M_\mathrm{env} + M_\mathrm{c}$, it is straightforward to show that this minimum occurs at $M_\mathrm{env} = \frac{1+\beta}{2\beta+3} M_\mathrm{c}$. A similar result is numerically shown in Fig. 4 of \citet{LeeEtal2014}. Their luminosity-minimum appears at a lower envelope mass, an indication of the importance of envelope self-gravity in models with $\gamma>4/3$.

In the following, we consider envelopes that are still well below the transition point to runaway accretion, such that $M_\mathrm{env} \ll M_\mathrm{p}$. This is a good description for all planets with metal-free envelopes that do not end their evolution as gas giants. To proceed, we establish the scaling between the energy released by cooling and the envelope mass. We start by considering the dominant energy term for low-mass planets, the thermal heat of the core. From Eq. \ref{eq:T_ad_xy}, we can approximate the temperature of the gas at the core-envelope boundary ($T_\mathrm{cg}$) as 
\begin{equation}\label{eq:T_cg}
    T_\mathrm{cg} \simeq T_\mathrm{disk} \frac{r'_\mathrm{B}}{r_\mathrm{c}} \propto M_\mathrm{p}.
\end{equation}
Since this temperature scales with the total mass, whose change is small in the low-mass regime, its relative change is minimal. Besides, the core also forms a temperature gradient and does not increase isothermally. We therefore do not include the core's energy in a planets embedded cooling.

Instead, the most important energy terms in the context of embedded cooling are those that scale with envelope mass and thus yield a significant time-derivative (luminosity). These are the gravitational and thermal energies of the envelope:
\begin{subequations}
\begin{align}
    E_\mathrm{env} &= \Phi_\mathrm{env} + U_\mathrm{env, th} \\
     &= -\int_{M_\mathrm{c}}^{M_\mathrm{p}}\frac{GM_\mathrm{c}}{r}dm + \int_{M_\mathrm{c}}^{M_\mathrm{p}} C_\mathrm{V} T_\mathrm{g}(m) dm \\
    &\simeq -\frac{G M_\mathrm{c} M_\mathrm{env}}{r_\mathrm{c}} {\left(\frac{r_\mathrm{c}}{r_\mathrm{rcb}}\right)}^\frac{3\gamma_\mathrm{xy}-4}{\gamma_\mathrm{xy}-1} \frac{3\gamma_\mathrm{xy}-4}{3-2\gamma_\mathrm{xy}}\frac{\gamma_\mathrm{xy}-1}{\gamma_\mathrm{xy}} \label{eq:E_env_XY} \\
    &\approx -1.09\frac{G M_\mathrm{c} M_\mathrm{env}}{r_\mathrm{c}} {\left(\frac{r_\mathrm{c}}{r_\mathrm{rcb}}\right)}^{0.78},
\end{align}
\end{subequations}
where we used $C_\mathrm{V} = \frac{1}{\gamma_\mathrm{xy}-1} \big(k_\mathrm{B}/\mu_\mathrm{xy}\big)$ and substituted the envelope mass through Eq. \ref{eq:M_env_XY} to simplify the result. Since $r_\mathrm{rcb}$ is proportional to the total mass (through the Bondi radius, see Eq. \ref{eq:r_rcb}), the energy in the low-mass limit scales as $-E_\mathrm{env} \propto M_\mathrm{env}$, such that $\dot{E}_\mathrm{env} \propto -\dot{M}_\mathrm{env}$. With this scaling relation, we can rewrite Eq. \ref{eq:L_scaling_xy} as a first-order ODE
\begin{equation}
    L = -\frac{dE_\mathrm{env}}{dt} = L_\mathrm{0} {\left(\frac{E_\mathrm{env}}{E_\mathrm{env,0}}\right)}^{-{(1+\beta)}}.
\end{equation}
Hence, the envelope energy and mass continue to evolve as
\begin{equation}\label{eq:u_t_metal-poor}
    \frac{E_\mathrm{env}}{E_\mathrm{env, 0}} = \frac{M_\mathrm{env}}{M_\mathrm{env, 0}} = {\left(1 + (2+\beta) \frac{t}{\tau_\mathrm{KH,0}} \right)}^\frac{1}{2+\beta},
\end{equation}
where $\tau_\mathrm{KH,0} = -E_\mathrm{env,0}/L_0$ is the Kelvin-Helmholtz timescale at $t_0 = 0$, defined as the start of phase III. Equation \ref{eq:u_t_metal-poor} shows that contracting metal-free planets roughly double their envelope mass after one Kelvin-Helmholtz timescale, consistent with previous works \citep[e.g.,][]{Lee2015, Lee2019}.

\subsection{Polluted envelope cooling inside a disk}\label{sect:enriched_cooling}
We will now show that post-accretion cooling proceeds much more slowly if a portion of the metals is contained in the envelope. The general derivation we perform is similar to that of the previous subsection, but modified due to the dominant presence of vapor in the inner envelope.

As before, we begin by determining the mass-scaling relation of the luminosity. We showed in Sect. \ref{sect:phase_II} that the total envelope mass can be approximated by that of the high-Z region and that it scales as $M_\mathrm{env} \propto \rho_\mathrm{vap} (\mu_\mathrm{g}/\mu_\mathrm{xy})^\frac{1}{\gamma_\mathrm{g}-1} M_\mathrm{p}^\frac{1}{\gamma_\mathrm{g}-1}$ (see Eq. \ref{eq:M_env_approx}), more steeply than a metal-free envelope. Again, this expression depends on $\rho_\mathrm{rcb}$, now through $\rho_\mathrm{vap}$ (see Eq. \ref{eq:rho_step}). We can eliminate this dependency with Eq. \ref{eq:rho_rcb_scaling} and write the luminosity scaling as
\begin{equation}\label{eq:L_scaling_z}
    \frac{L}{L_0} = {\left(\frac{M_\mathrm{env}}{M_\mathrm{env,0}}\right)}^{-(1+\beta)} {\left(\frac{M_\mathrm{p}}{M_\mathrm{p,0}}\right)}^\frac{\gamma_\mathrm{g}+\beta}{\gamma_\mathrm{g}-1} {\left(\frac{\mu_\mathrm{g}}{\mu_\mathrm{g,0}}\right)}^\frac{\gamma_\mathrm{g}(1+\beta)}{\gamma_\mathrm{g}-1}.
\end{equation}
Comparison with Eq. \ref{eq:L_scaling_xy} reveals that it contains an additional term involving the mean molecular weight that describes the dilution of the envelope. Any contraction leads to the inflow of metal-poor nebular gas, causing $\mu_\mathrm{g}$ to decrease from its initial value of $\mu_\mathrm{g,0}$. This dilution essentially helps to regulate the planet's further contraction by lowering the density in the region where the majority of the envelope mass is contained. The planet consequently accretes nebular gas more slowly than it would if its composition remained unchanged.

\subsubsection{The energy of polluted envelopes}
Similar to the case of metal-free envelopes, the terms that evolve during embedded cooling are the gravitational and thermal components. Because the envelope does not decrease in temperature while the envelope increases in mass, condensation and latent heat release are unimportant to the embedded cooling process. The gravitational term is slightly complicated compared to the metal-free case due to the non-negligible importance of self-gravity in polluted envelopes and can be written as
\begin{subequations}
\begin{align}
    \Phi_\mathrm{env} &= -\int_{M_\mathrm{c}}^{M_\mathrm{p}}\frac{G\big(M_\mathrm{c} + M_\mathrm{env}(r)\big)}{r}dm \label{eq:phi_env_int}\\
    &\simeq -\frac{G M_\mathrm{c} M_\mathrm{env}}{r_\mathrm{c}} \frac{4-3\gamma_\mathrm{g}}{3-2\gamma_\mathrm{g}} \left(1 + \frac{4-3\gamma_\mathrm{g}}{7-5\gamma_\mathrm{g}} \frac{M_\mathrm{env}}{M_\mathrm{c}}\right),
\end{align}
\end{subequations}
where we substituted the envelope mass expression from Eq. \ref{eq:M_env_int} and expanded this into the integral of Eq. \ref{eq:phi_env_int}. We then applied the relevant limit of $r_\mathrm{vap} \gg r_c$ to simplify the result. We also evaluate the thermal term and find that it is similar to the expression for metal-free envelopes, but without the assumption of a dominating core mass:
\begin{subequations}
\begin{align}
    U_\mathrm{env, th} &= \int_{M_\mathrm{c}}^{M_\mathrm{p}} C_\mathrm{V} T_\mathrm{g} dm \\
    &\simeq \frac{G M_\mathrm{env}}{r_\mathrm{c}} \frac{4-3\gamma_\mathrm{g}}{\gamma_\mathrm{g}(3-2\gamma_\mathrm{g})}\left(M_\mathrm{env}+M_\mathrm{c}\right),
\end{align}
\end{subequations}
where we used $C_\mathrm{V} = \frac{1}{\gamma_\mathrm{g}-1} \big(k_\mathrm{B}/\mu_\mathrm{g}\big)$. Together with the gravitational potential, this yields the total envelope energy
\begin{subequations}
\begin{align}
    E_\mathrm{env} &= \Phi_\mathrm{env} + U_\mathrm{env, th}\\
    &\simeq -\frac{G M_\mathrm{c} M_\mathrm{env}}{r_\mathrm{c}}\frac{4-3\gamma_\mathrm{g}}{3 - 2\gamma_\mathrm{g}} \left(\frac{\gamma_\mathrm{g}-1}{\gamma_\mathrm{g}} + \left(\frac{4-3\gamma_\mathrm{g}}{7-5\gamma_\mathrm{g}}-\frac{1}{\gamma_\mathrm{g}}\right) \frac{M_\mathrm{env}}{M_\mathrm{c}}\right) \label{eq:E_env_enriched} \\
    &\approx -\frac{G M_\mathrm{c} M_\mathrm{env}}{r_\mathrm{c}} \left(0.1 - 0.23 \frac{M_\mathrm{env}}{M_\mathrm{c}}\right).
\end{align}
\end{subequations}
Eq. \ref{eq:E_env_enriched} consists of two terms: a negative scaling with $M_\mathrm{env} M_\mathrm{c}$ that accounts for the gravitational potential of the core and a positive scaling with $M_\mathrm{env}^2$ that encapsulates the effects of self-gravity. The importance of this second term is a characteristic of polluted envelopes, as only a portion of the metals are located at the core. As long as the envelope is small in mass relative to the core, the expression is very similar to that of metal-free envelopes (Eq. \ref{eq:E_env_XY}). But polluted planets can be characterized by more massive envelopes, where the self-gravity term of the envelope energy becomes first-order in their energy budget. In these cases, Eq. \ref{eq:E_env_enriched} turns positive when $M_\mathrm{env} \gtrsim 0.43 M_\mathrm{c}$. This result is analogous to the energy calculation of an ideal gas stellar polytrope with $\gamma < \frac{4}{3}$ that also yields a positive energy. The only way to actually reach such an energy, however, is if the net luminosity is negative when integrated over the full accretion time. In other words: the envelope is not bound. Such a scenario is clearly non-physical for a number of reasons, so Eq. \ref{eq:E_env_enriched} cannot be an accurate representation of a massive, polluted planet's energy. 

Still, a few meaningful observations can be made from it. First, it shows that the assumption of uniform mixing of the interior region breaks down in the high-mass limit if $\gamma_\mathrm{g}<4/3$, as there is \textrm{eventually} insufficient energy to transport the material upwards. Secondly, Eq. \ref{eq:E_env_enriched} indicates that the typical assumption of a global luminosity in formation and internal evolution models (including ours) is unfeasible in the high-mass polluted regime. Dealing with these complications will be a challenge for all new interior evolution models that wish to study high-mass polluted envelopes in more detail.

\subsubsection{Estimating gas accretion}
Due to the reasons outlined in the previous paragraph, our analytical model is ill-suited to accurately calculate the absolute contraction timescale of a massive, polluted envelope in a disk. Without considering the thermal heat, a polluted envelope contains more energy due to its scaling with $1/r_\mathrm{c}$ instead of $1/r_\mathrm{rcb}$, as most of the gas mass is located at its core. This would indicate a greater Kelvin-Helmholtz timescale than a metal-poor envelope. However, the assumption of perfect mixing in the interior yields an overestimation of the thermal energy, that overwhelms even this larger gravitational potential. In reality, a polluted planet with $\gamma_\mathrm{g}<\frac{4}{3}$ does not have sufficient energy to mix its material uniformly over its high-Z region, and the distribution would be more skewed towards the core. Still, its interior would be bound and thus not contribute positively to the energy balance. Consequently, polluted envelopes may not have significantly greater kelvin-Helmholtz timescales, but they will also not be far smaller than those of metal-poor envelopes, who contain their mass as the radiative-convective boundary and are already at the lower energy limit.

It is, therefore, still possible to distill some insights into their cooling by considering normalized equations in terms of $\tau=t/\tau_\mathrm{KH,0}$. For the sake of simplifying notation and increased clarity, we also introduce the variable $\epsilon = M_\mathrm{xy}/M_\mathrm{z}$. As long as the planet has accreted quantities of nebular gas that are small compared to the mass in metals (core and vapor) and is thus not close to entering runaway, the relevant regime to consider is the limit $\epsilon \ll 1$, where we can write (see appendix A):
\begin{equation}\label{eq:cooling_polluted_lowmass}
    \frac{\epsilon}{\epsilon_0} = \frac{M_\mathrm{xy}}{M_\mathrm{xy,0}} = {\left(1 + \frac{f_\mathrm{z}}{\epsilon_0} \frac{(2+\beta)\gamma_\mathrm{g}-1}{\gamma_\mathrm{g}-1} \frac{t}{\tau_\mathrm{KH,0}} \right)}^\frac{\gamma_\mathrm{g}-1}{(2+\beta)\gamma_\mathrm{g}-1}.
\end{equation}
Equation \ref{eq:cooling_polluted_lowmass} has the same form as Eq. \ref{eq:u_t_metal-poor}, which we derived for the cooling of unpolluted planets. The difference is that the constant that appears in front of $t/\tau_\mathrm{KH,0}$ and inversely in the exponent, is much greater for polluted envelopes. This shows that while these planets initially experience more rapid contraction during embedded cooling, it slows down over time. Physically, this is due to the previously mentioned dilution of the interior region.

\begin{figure}[t!] 
\centering
\includegraphics[width=\hsize]{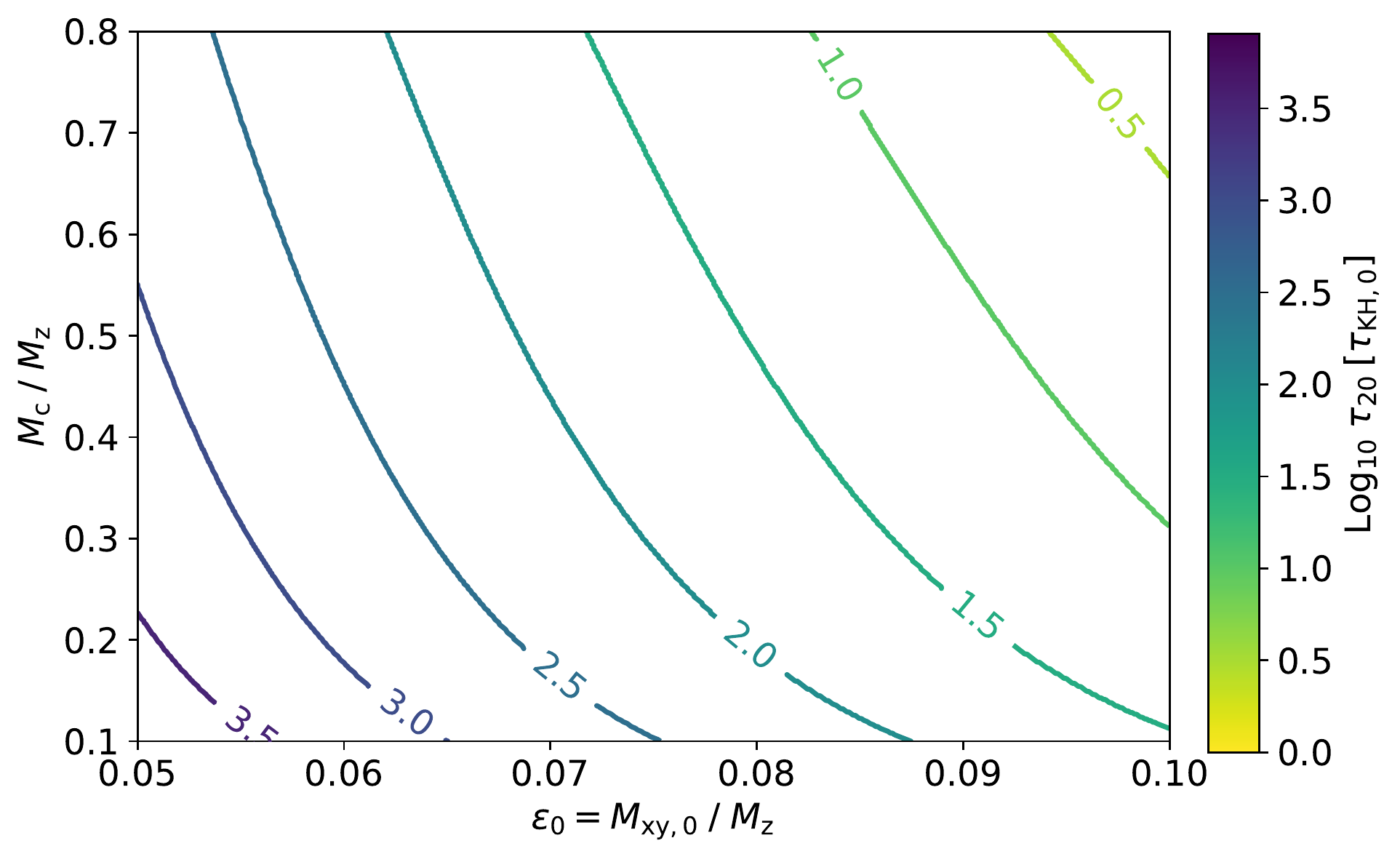} 
\caption{Time $\tau_{20}$ (in $\tau_\mathrm{KH,0}$) required for polluted envelopes with initial hydrogen-helium mass fraction $\epsilon_0$ to accrete 20\% nebular gas. The figure shows that heavily polluted planets accrete gas more slowly in terms of their Kelvin-Helmholtz timescales, but that the initial nebular gas fraction is the main determinant.}\label{fig:cooling_20}
\end{figure}

We visualize this result in Fig. \ref{fig:cooling_20}, which plots the time required for a polluted planet with hydrogen-helium mass fraction $\epsilon_0$ to accrete up to a mass fraction of 20\% nebular gas ($\epsilon(\tau_{20}) = 0.20$). It indicates that contraction proceeds more slowly for planets with small cores, as they are more severely polluted and are more effectively stabilized by dilution. The figure also shows that the initial nebular gas fraction is the main factor in determining the gas accretion rate.

\begin{figure}[t!] 
\centering
\includegraphics[width=\hsize]{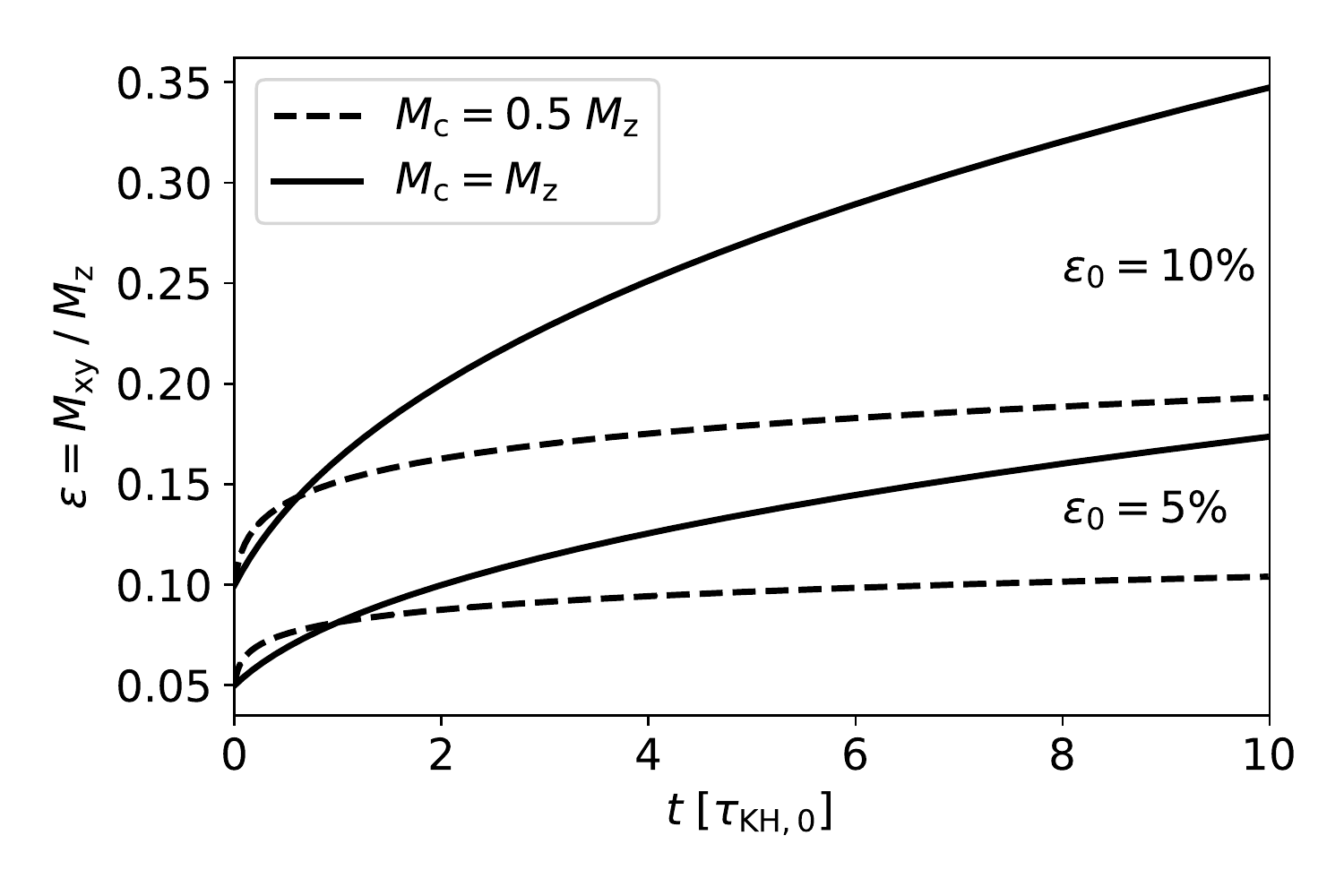} 
\caption{Embedded cooling comparison between metal-free ($M_\mathrm{c}=M_\mathrm{z}$, solid lines) and polluted ($M_\mathrm{c}=0.5 \; M_\mathrm{z}$, dashed lines) grain-free envelopes. The lower and upper lines show initial hydrogen-helium mass fractions of 5\% and 10\%, respectively. It is clear that while polluted envelopes initially accrete significant amounts of nebular gas, their contraction slows down considerably due to the dilution of the envelope. }\label{fig:cooling_analytical}
\end{figure}

We also offer comparison with metal-free envelopes in Fig. \ref{fig:cooling_analytical}, where we plot the embedded cooling of a polluted and metal-poor envelope, that contain equal hydrogen-helium mass fractions of 5 \% (but have different metal masses). The figure again shows that the presence of high-Z vapor inhibits the contraction of the planet, causing it to require a greater number of Kelvin-Helmholtz timescales to double in mass. This important trend did not appear in previous works that considered embedded cooling \citep[e.g.,][]{Lee2015} because we do not assume that $Z$ remains constant during cooling and instead allow the composition to change with the accretion of additional nebular gas. Note that the figure does not imply that polluted planets with a given metal mass accrete less nebular gas than their metal-free counterparts, as we do not calculate the absolute cooling timescale. In fact, enriched envelopes will likely contract more rapidly when they have a similar $M_\mathrm{z}$, as they contain significantly more nebular gas already ($\epsilon_\mathrm{0}$ is larger). Rather, we find that polluted envelopes require a greater number of Kelvin-Helmholtz timescales than metal-free envelopes with the same hydrogen-helium mass fraction (same $\epsilon_\mathrm{0}$).

\section{Phase IV: Cooling post disk dissipation (indirect core growth)}\label{sect:phase_IV}
Ultimately, after several Myr of accretion, the disk dissipates and its planets become exposed to the vacuum of space. Without any surrounding gas left to to accrete, they can only contract and their radii shrink while their interiors cool down. This is a fundamental difference with the embedded cooling described in the previous section, where a planet's size still effectively increases due to the inflow of nebular gas. In our model, the temperature decrease provides yet another important distinction with embedded cooling, as the increased saturation will eventually force the high-Z component to condense out and sediment. This will cause the mass of the core to increase again, a process that we refer to here as indirect core growth (to contrast with direct core growth in Phase I). Indirect core growth potentially allows for the formation of larger, super-Earth sized cores, with masses up to the planet's lifetime integrated high-Z flux. 

\subsection{Mass-loss during cooling}\label{sect:massloss}
\begin{figure*}[t!] 
\centering
\includegraphics[width=\hsize]{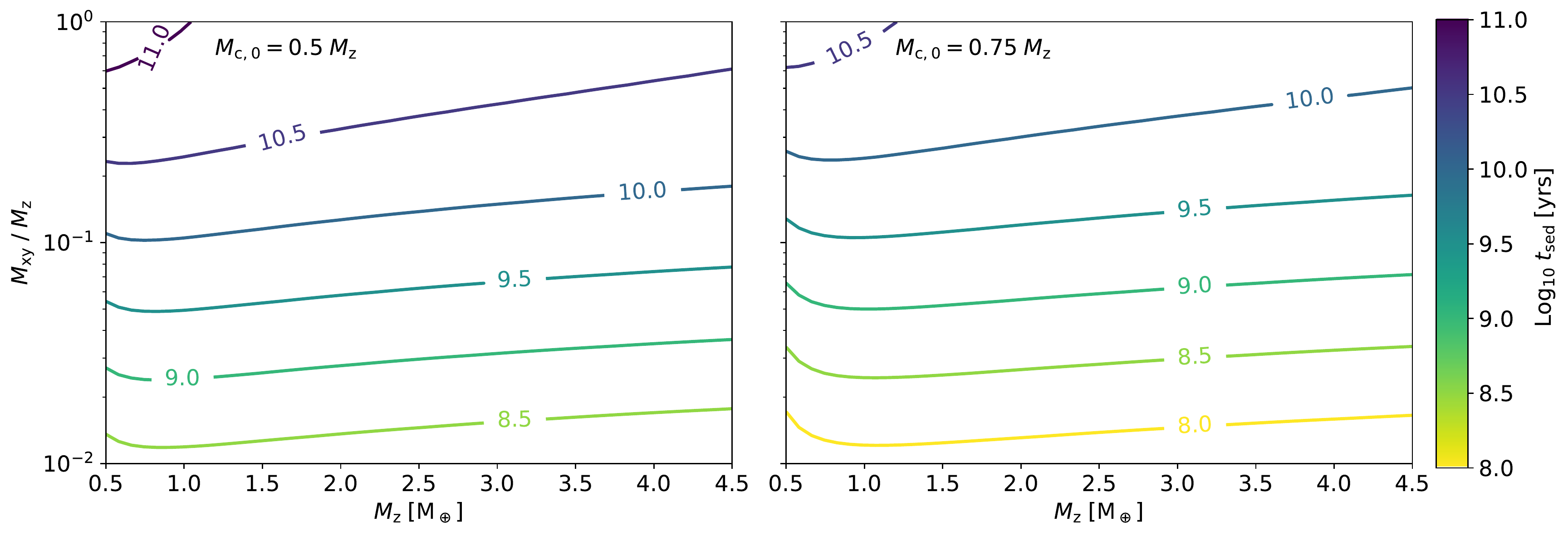} 
\caption{Time $t_\mathrm{sed}$ that planets are required spend in isolation in order to sediment all the vapor contained in their high-Z regions. The values correspond to a grain-free planet at 0.1 AU, calculated with our defeault parameters (see Table \ref{table:default_parameters}). The left and right panels assume initial core mass fractions of 0.5 and 0.75, respectively.  }\label{fig:cooling_sed}
\end{figure*}
Newly exposed planets also become vulnerable to two distinct mass-loss mechanisms: photo-evaporation and vacuum-induced outflows. The first process refers to the idea that high-energy photons can efficiently remove molecules from the envelope \citep[e.g.,][]{Baraffe2004, Hubbard2007, Murray2009, Valencia2010, Lopez2013}, or even from the planet's core if it is located sufficiently close-in \citep{Becker2013}. It is a well-studied phenomenon by now and has been suggested by \citet{OwenWu2017} to explain the bimordial distribution in exoplanetary radii \citep{Fulton2017,Fulton2018}, as it is only effective at removing low-mass envelopes. There is no reason to suggest that photo-evaporation is influenced by the degree of envelope pollution, as the gas is removed from the metal-free outer layers from the envelope.

Vacuum-induced outflows, in contrast, refer to the basic idea that the loss of disk-pressure induces a steady stream of gas leaving the planet's envelope, supported by the release of energy that accompanies it \citep{Ikoma2012, Ginzburg2016, Ginzburg2018}. The associated mass-loss rate can be very large, but is heavily dependent on the flow morphology. If an outflow is isothermal, its rate is given by \citep{Parker1958, Owen2016}
\begin{equation}
    \dot{M}_\mathrm{xy} = 4\pi r_\mathrm{B}^2\textit{M}_\mathrm{p} \frac{P_\mathrm{surf}}{c_\mathrm{s}},
\end{equation}
where $\textit{M}_\mathrm{p}$ is the Mach number and $P_\mathrm{surf} \approx g/\kappa_\mathrm{surf}$ is the planet's photospheric surface density. As a planet shrinks and arrives in the limit where $r_\mathrm{p} \ll r_\mathrm{B}$, the flow's Mach number begins to decline exponentially and eventually drops to a negligible value when $r_\mathrm{p} \approx 0.1 r_\mathrm{B}$ \citep{Cranmer2004}. This means that the outflow is always self-limited and cannot remove entire envelopes on its own, without the aid of photo-evaporation. Still, the resultant mass-loss can potentially be significant and is worth evaluating.

In the case of metal-free envelopes, the energy that powers the molecules leaving the atmosphere is hypothesized to come from the release of gravitational energy during contraction. From Eq. \ref{eq:E_env_XY}, it is evident that their energy scales as $-E_\mathrm{env} \propto M_\mathrm{env}/r_\mathrm{rcb}$. As the planet contracts, $r_\mathrm{rcb}$ drops significantly and $-E_\mathrm{env}$ thus increases. However, this mechanism operates differently for polluted envelopes, whose energy instead scales like $E_\mathrm{env} \propto M_\mathrm{env}/r_\mathrm{c}$ (see Eq. \ref{eq:E_env_enriched}). In this case, the only way to release significant amounts of energy is for the planet to cool sufficiently, so that its interior vapor region begins to sediment, allowing for the release of the vast amounts of energy stored in the form of latent heat and gravitational potential. 

We aim to estimate whether this sedimentation occurs before the planet has shrunk to $r_\mathrm{p} \approx 0.1 r_\mathrm{B}$, so that its release can lead to significant mass-loss before it is lost to radiation. We begin by considering the photospheric surface density $\rho_\mathrm{ph}$. Since the planet's surface temperature due to irradiation is comparable to the previous disk temperature, this can simply be written as 
\begin{equation}
    \rho_\mathrm{ph} = \frac{r_\mathrm{B}}{\kappa_\mathrm{ph}r_\mathrm{p}^2}.
\end{equation}
We can roughly estimate the condition for sedimentation to be when the gas temperature at the core boundary is insufficient for significantly vapor absorption, so that $T_\mathrm{cg} = T_\mathrm{vap}$. With Eq. \ref{eq:rho_ad_xy}, this yields
\begin{equation}\label{eq:r_rcb_sed}
    \frac{r_\mathrm{rcb}}{r_\mathrm{c}} = \frac{1}{1-\big(T_\mathrm{vap}/T_\mathrm{rcb}\big) \big(r_\mathrm{c}/r'_\mathrm{B}\big)} \approx \mathrm{O}(1).
\end{equation}
Determining $T_\mathrm{rcb}$ for an irradiated and exposed planet requires evaluating the radiative transport in the planet's atmosphere \citep[e.g.,][]{Guillot2010}. In the case of a heavily contracted planet, though, the resulting temperature will not be increased much by the internal luminosity. We therefore approximate it as the temperature that characterized the disk when it was still present, set by the star's irradiation ($T_\mathrm{rcb} \approx T_\mathrm{ph} \approx T_\mathrm{disk}$). In any case, the order-of-magnitude evaluation of Eq. \ref{eq:r_rcb_sed} is unaffected by this choice, since $r'_\mathrm{B} \ll r_\mathrm{c}$.

Finally, we can combine the previously made estimates to approximate the effective radius $r_\mathrm{p}$ during indirect core growth. Because $r_\mathrm{p}$ and $r_\mathrm{rcb}$ are connected by a growing radiative region, they obey Eq. \ref{eq:rho_rad} with $r_\mathrm{out} = \ r_\mathrm{p}$ and $\rho_\mathrm{disk} \rightarrow \rho_\mathrm{ph}$. When worked out, this yields
\begin{equation}
    \frac{r_\mathrm{p}}{r_\mathrm{rcb}} = \frac{1}{1+\mathrm{log}\big(\rho_\mathrm{rcb}/\rho_\mathrm{ph}\big) \big(r_\mathrm{rcb}/r_\mathrm{B}\big)}.
\end{equation}
Since $\mathrm{log}\big(\rho_\mathrm{rcb}/\rho_\mathrm{ph}\big)$ can never grow very large as a logarithm and because $r_\mathrm{rcb} \ll r_\mathrm{B}$ after a significant period of cooling, we can comfortably estimate $r_\mathrm{p}\approx r_\mathrm{rcb}<0.1 r_\mathrm{B}$ during indirect core growth. So regardless of how much energy is stored in the high-Z region, it is only liberated late in the planet's cooling, after the planet has contracted to $r_p \ll 0.1 r_B$. Therefore, we expect vacuum induced outflows to be suppressed in vapor-rich planets.

\subsection{Connection to current sub-Neptune interiors}\label{sect:highZloss}
In order to make predictions on the current interiors of observed exoplanets, we need to estimate how long it takes for their high-Z region to completely sediment. We showed in the last subsection that $r_\mathrm{rcb} \approx r_\mathrm{c}$ during indirect core growth and that the envelope mass is thus preserved, unless it is removed by high energy photons. If we do not consider photo-evaporation, a slightly modified version of Eq. \ref{eq:M_env_XY} (that does not assume $r_\mathrm{rcb} \gg r_\mathrm{c}$) provides a good approximation for the envelope mass after the vapor has sedimented:
\begin{equation}
    M_\mathrm{xy} = 
    4\pi\frac{\gamma_\mathrm{xy}-1}{3\mathrm{\gamma_\mathrm{xy}}-4} \rho_\mathrm{rcb} {r'}_\mathrm{B}^\frac{1}{\gamma_\mathrm{xy}-1} \left(r_\mathrm{rcb}^\frac{3\gamma_\mathrm{xy}-4}{\gamma_\mathrm{xy}-1} - r_\mathrm{c}^\frac{3\gamma_\mathrm{xy}-4}{\gamma_\mathrm{xy}-1}\right). \label{eq:M_env_XY_general}
\end{equation}
If we parameterize $M_\mathrm{xy}$, the above equation directly yields $\rho_\mathrm{rcb}$ at the end of indirect core growth:
\begin{subequations}
\begin{align}
    \frac{\rho_\mathrm{rcb}}{\rho_\mathrm{c}} &= \frac{3\gamma_\mathrm{xy}-4}{3(\gamma_\mathrm{xy}-1)} {\left(\frac{r_\mathrm{c}}{r'_\mathrm{B}}\right)}^\frac{1}{\gamma_\mathrm{xy}-1} \frac{M_\mathrm{xy}}{M_\mathrm{c}} {\left({\left(\frac{r_\mathrm{rcb}}{r_\mathrm{c}}\right)}^\frac{3\gamma_\mathrm{xy}-4}{\gamma_\mathrm{xy}-1}-1\right)}^{-1} \\
    &\simeq 
    \frac{1}{3} {\left(\frac{r_\mathrm{c}}{r'_\mathrm{B}}\right)}^\frac{2-\gamma_\mathrm{xy}}{\gamma_\mathrm{xy}-1} \frac{T_\mathrm{rcb}}{T_\mathrm{vap}} \frac{M_\mathrm{xy}}{M_\mathrm{c}}, \label{eq:rho_rcb_sed}
\end{align}
\end{subequations}
where we substituted Eq. \ref{eq:r_rcb_sed} in the second step. We can use Eq. \ref{eq:rho_rcb_sed} to estimate the resultant luminosity $L_\mathrm{sed}$ from Eq. \ref{eq:L_rad_diff} and find:
\begin{equation}
    L_\mathrm{sed} = \frac{64\pi\sigma T_\mathrm{rcb}^3 T_\mathrm{vap}}{\kappa_\mathrm{rcb} \rho_\mathrm{c}} r_\mathrm{c}^{-\frac{2-\gamma_\mathrm{xy}}{\gamma_\mathrm{xy}-1}}
    {r'}_\mathrm{B}^\frac{1}{\gamma_\mathrm{xy}-1} \frac{M_\mathrm{c}}{M_\mathrm{xy}}. \\
\end{equation}
If we substitute the opacity scaling for a grain-free envelope, this yields
\begin{equation}
    L_\mathrm{sed, gf} = \frac{64\pi\sigma T_\mathrm{rcb}^{-\frac{2}{3}} T_\mathrm{vap}^\frac{5}{3} r'_\mathrm{B}}{3^\frac{2}{3} \kappa_\mathrm{cst} \rho_\mathrm{c}^\frac{5}{3}}
    {\left(\frac{M_\mathrm{xy}}{M_\mathrm{c}}\right)}^{-\frac{5}{3}}
    {\left(\frac{r_\mathrm{c}}{r'_\mathrm{B}}\right)}^{-\frac{5(2-\gamma_\mathrm{xy})}{3(\gamma_\mathrm{xy}-1)}}.
\end{equation}
The temperature scaling of $\kappa_\mathrm{rcb, gf}$ implies that the dependence on $T_\mathrm{rcb}$, the largest simplification in this calculation, largely cancels out.

The final step is to estimate the total energy reserve of the high-Z region $E_\mathrm{sed}$, composed of the latent heat and gravitational terms:
\begin{subequations}
\begin{align}\label{eq:E_cool}
    E_\mathrm{sed} &= U_\mathrm{latent} + \Delta\Phi_\mathrm{c} \\
    &\approx \left(M_\mathrm{z} - M_\mathrm{c}\right) u_\mathrm{L} + \frac{3}{5}\left(\frac{M_\mathrm{z}^2}{R_\mathrm{c}} - \frac{M_\mathrm{c}^2}{r_\mathrm{c}} \right),
\end{align}
\end{subequations}
where $R_\mathrm{c}$ is the core's radius after indirect growth has proceeded. We plot the resultant time ($t_\mathrm{sed} = -E_\mathrm{sed}/L_\mathrm{sed}$) required to sediment all the vapor contained in the high-Z region in Fig. \ref{fig:cooling_sed}. The figure shows that only small planets that recently formed a vapor region can cool within a few Gyr. More massive planets with envelopes that contain a few percent of hydrogen-helium gas by mass take about 10 Gyr to sediment their high-Z vapor. The long timescales we find are the result of the planet being heavily contracted, the same effect that leads to long core-cooling timescales in unpolluted super-Earths \citep{Vazan2018}.

\section{Discussion}\label{sect:discussion}
The trends that we have identified regarding the evolution of polluted envelopes are especially relevant to the formation of super-Earths and sub-Neptunes and can potentially help to explain their observed distribution. Instead of calculating how the critical core mass is modified by pollution, as has been done for enriched envelopes in previous works \citep{Stevenson1982, Wuchterl1993, Venturini2015, Venturini2016}, we show that the more natural criterion for runaway accretion is a limit on the total of the accreted solids (core + vapor), which we refer to as the critical metal mass. Crucially, our derived expressions for this quantity show an inverse scaling with $T_\mathrm{vap}$, the temperature at the outer boundary of the vapor region that indicates the volatility of the high-Z material. More volatile constituents like water ice evaporate at lower temperatures and thus lead to a larger vapor region. Their accretion also results in smaller core masses, another quantity that we find is positively linked to the critical metal mass. Both a small core and a large vapor region increase the amount of nebular gas that is drawn in by the accreted solids and thus accelerate the onset of runaway accretion. Planets that form at greater orbital separations, outside the ice-line, accrete pebbles and planetesimals that contain a larger fraction of volatile ices. Our calculation therefore indicates that far-out planets reach runaway at lower masses, compared to those that form close-in. If planets accrete similar amounts of solids at any disk location, the result is that the abundance of super-Earths and sub-Neptunes relative to gas giants becomes naturally skewed towards the inner disk.

Our finding that pollution enhances the accretion of nebular gas is in agreement with \citet{Stevenson1982, Wuchterl1993, Venturini2015, Venturini2016} but is contrary to the result by \citet{Bodenheimer2018} and highlights a key difference with this work. Because we assume a uniformly mixed interior region, the presence of vapor actively helps to suck in nebular gas. It simultaneously increases the mean molecular weight and decreases the adiabatic index, thus raising the envelope density. In contrast, \citet{Bodenheimer2018} assume that the envelope does not compositionally mix, and hence they find an interior envelope (referred to as an "outer core") that consists entirely of silicates. Because this region has a lower density than the central core, the surrounding metal-poor envelope in their model accretes nebular gas at a slower rate than it would if all the metals were to sediment. The poorly constrained diffusive constant in hot planetary interiors currently makes it difficult to discern what scenario is more realistic.

The subsequent phase of embedded cooling is equally important in explaining the observed planetary distribution. Unless solids accretion is fine-tuned to yield predominantly super-Earth masses within the disk lifetime, many planets that end up in the intermediate mass-range must experience an extended period of cooling while the disk is still present, for instance due to a planet further out reaching the pebble isolation mass, restricting the pebble flow to the inner disk \citep{Morbidelli2012, Lambrechts2014, Morbidelli2015}. This necessity of an embedded cooling phase is problematic for current formation models, as metal-free envelopes contract rapidly unless they have a large opacity and soon thus reach runaway gas accretion if they initiate their cooling with even a small percentage of nebular gas by mass \citep{LeeEtal2014}. While we cannot make predictions on the absolute contraction timescale of polluted planets, we find that they require many more Kelvin-Helmholtz timescales to double their disk gas content, compared to unpolluted planets with the same hydrogen-helium mass fraction. Physically, this is due to the stabilizing dilution of the interior. This connection can help explain the preservation of the intermediate-mass planets that contain a few percent primordial nebular gas, a prerequisite in explaining the observed sub-Neptune population. 

The final phase of any planet is the cooling on Gyr timescales that follows the dissipation of the primordial disk. We find that in our model, the inverse energy scaling of polluted planets with the core radius, instead of the total radius, makes vacuum-induced outflows much less efficient. Physically, this is because we characterize their hot, polluted interiors with $\gamma_\mathrm{g} < \frac{4}{3}$, such that most of their envelope mass is contained close to their cores, instead of at the radiative-convective boundary. The efficiency of photo-evaporation remains unchanged, however, and thus becomes the primary channel for mass-loss. Our estimation of the time required to sediment the high-Z region indicates timescales of several Gyr. This indicates that the observed sub-Neptunes likely do not have cores that hold their complete metal contents, but instead have a more complex interior \citep{Vazan2016}. Similarly, Jupiter only contains a portion of its high-Z mass in its central core \citep{Wahl2017, Helled2017, Helled2018}. In our model, this is a natural consequence of the preservation of its inner high-Z regions and is a remnant of the planet's accretion phase, rather the result of core erosion.

The main caveat to our calculations is that we assume a perfectly mixed interior and use an ideal equation of state to retain analytical expressions. We find in Sect. \ref{sect:runaway_analytical} that an enriched ideal gas, together with a small adiabatic index, can result in nonphysically large gas densities in the envelope's deep interior. Similarly, we find in Sect. \ref{sect:enriched_cooling} that the energy of these envelopes turns positive when the envelope mass is comparable to that of the core. These instances indicate that the assumption of uniform compositional mixing is unlikely to be a good approximation. As the high-Z region expands, the vapor that it contains has to be moved upwards. The energy available for upward motion ultimately disappears when the absorption of accreted solids occurs close to the region's outer edge. Accounting for this effect would result in a greater concentration of the high-Z material around the core, more in line with results from \citet{Lozovsky2017, Bodenheimer2018}. These limitations imply that we can only model the embedded cooling phase (III) in terms of scaling relations, where the evolution is expressed in terms of the KH-timescale, and cannot make quantitative predictions. Our results therefore show the need for numerical simulations that do not make the assumption of a uniform interior composition with ideal gases, enabling it to refine our approximation of the critical metal mass and to calculate cooling on absolute timescales. Such simulations are numerically difficult to perform in a stable manner, and require a more involved approach than the simple numerical comparison we provided in this paper. We will pursue this next step in a subsequent work.

Finally, our model makes the assumption that the high-Z region is characterized by an adiabatic index $\gamma_\mathrm{g} < \frac{4}{3}$, and that envelopes nevertheless remain stable against dynamical collapse during their formation. Both assumptions are uncertain, and require further investigation. Determining the value of the adiabatic index in the deep interior is an especially complex task that at minimum demands an equation-of-state that accounts for chemical reactions, as well as the rate of dissociation and ionization. The possibility of collapse was already mentioned by \citet{Hori_Ikoma_2011}, but has not been addressed in any study since. If polluted planets are indeed unstable, this would fundamentally change their evolution and likely invalidate our results. The question is whether the stability criterion of $\gamma>4/3$, which is derived for self-gravitating spheres (i.e. stars or clouds), can be generalized to planets, which are different in the sense that they contain a rigid core at their center. \citet{Wuchterl1990} performed an approximate analytical calculation assuming homologous oscillations and found that the presence of a core can have a significant stabilizing influence. Its relation to polluted envelopes emphasizes the importance of understanding planetary envelope stability more deeply and provides good motivation for follow-up work.
 
\section{Conclusions}\label{sect:conclusion}
Planets that form by the accretion of solids naturally experience envelope pollution, as impactors sublimate before they reach the core (see Fig. \ref{fig:impact_regimes} in the appendix). We have described the thermal evolution of these planets in the context of a simple analytical model, whose distinguishing feature is a layer of high-Z vapor with uniform composition and $\gamma_\mathrm{g}<\frac{4}{3}$, sandwiched between the core and the usual (metal-poor) envelope (see Figs. \ref{fig:zones} and \ref{fig:interior_conditions}). Our main findings are:

\begin{itemize}
  \item[$\bullet$] The evolution of polluted planets can naturally be described by four distinct phases (Fig. 1). I. Direct core growth: the core of the planet grows proportionally with the accretion of solids at close to full efficiency. II. Polluted envelope growth: solids completely vaporize in the hot, inner envelope, which becomes metal-rich and dominates its mass. III. Embedded cooling: solid accretion terminates and envelope growth occurs by virtue of nebular gas accretion. IV. Indirect core growth: after disk dispersal, protracted cooling could result in a renewed rainout of the vapor component, enhancing the core mass.
  \item[$\bullet$] We analytically derived a criterion for the runaway of polluted envelopes in phase II (Eqs. \ref{eq:M_crit_analytical_2}, \ref{eq:M_crit_analytical_3}). Its onset begins when a planet's total amount of metals (core + vapor) exceeds the stable limit that we refer to as the critical metal mass. This criterion naturally supersedes the critical core mass for planets with metal-free envelopes. The critical metal mass is typically smaller than the latter and scales inversely with the volatility of the vapor species.
  \item[$\bullet$] In phase III, polluted planets with equal hydrogen-helium mass fractions require more Kelvin-Helmholtz timescales to increase their mass, compared to unpolluted planets where accreted solids are assumed to sediment to the core. We derive a scaling relation to describe this embedded cooling phase (Eq. \ref{eq:L_scaling_z}) and identify the interior compositional dilution as a self-limiting factor in gas accretion.
  \item[$\bullet$] In our model, vacuum-induced outflows from polluted envelopes are an ineffective mass-loss process after the disk has dissipated (phase IV). The energy of polluted envelopes scales with the core radius, instead of the planet's total radius. This means that these planets do not release significant amounts of energy until the high-Z vapor sediments, at which point the planet is too contracted to shed mass efficiently. 
  \item[$\bullet$] The sedimentation of the vapor is sufficiently slow that it is possible for planets to retain part of their high-Z regions after several Gyr, as long as they conserve their primordial envelopes.
\end{itemize}

\section*{Acknowledgements}
\tiny{We are especially grateful to Allona Vazan, Carsten Dominik and Dave Stevenson for the insightful conversations we had on this topic. We also thank an anonymous referee for the comments that significantly improved the paper's clarity. Part of the work presented here is based on discussions conducted at the ISSI Ice Giants Meeting in Bern, 2019. C.W.O. is supported by The Netherlands Organization for Scientific Research (NWO; VIDI project 639.042.422).}

\bibliographystyle{aa}
\bibliography{analytical}
%
\begin{appendix}

\section{Derivation of Eq. \ref{eq:cooling_polluted_lowmass}}\label{Appendix_A}
\normalsize{
We begin by rewriting the changing mean molecular weight as a ratio of the envelope and accreted nebular gas mass:
\begin{equation}
    \frac{\mu_\mathrm{g}}{\mu_\mathrm{g,0}} = \frac{M_\mathrm{xy,0}}{M_\mathrm{xy}}\frac{M_\mathrm{env}}{M_\mathrm{env,0}}.
\end{equation}
If we substitute the above into Eq. \ref{eq:L_scaling_z}, this yields
\begin{equation}\label{eq:L_scaling_new}
    \frac{L}{L_0} = {\left(\frac{M_\mathrm{env}}{M_\mathrm{env,0}}\right)}^{\frac{1+\beta}{\gamma_\mathrm{g}-1}} {\left(\frac{M_\mathrm{p}}{M_\mathrm{p,0}}\right)}^\frac{\gamma_\mathrm{g}+\beta}{\gamma_\mathrm{g}-1} {\left(\frac{M_\mathrm{xy}}{M_\mathrm{xy,0}}\right)}^{-\frac{\gamma_\mathrm{g}(1+\beta)}{\gamma_\mathrm{g}-1}}.
\end{equation}
In order to link this luminosity to mass accretion, we note that the planet accretes only nebular gas during its embedded cooling phase and that both $M_\mathrm{z}$ and $M_\mathrm{c}$ remain constant. This means that we can approximate in the low-mass limit ($\epsilon = M_\mathrm{xy}/M_\mathrm{z} \ll 1$) that
\begin{subequations}
\begin{align}
    \frac{\dot{M}_\mathrm{env}}{M_\mathrm{env,0}} &= \frac{\dot{M}_\mathrm{xy}}{M_\mathrm{z}-M_\mathrm{c}+M_\mathrm{xy,0}} \simeq \dot{\epsilon} f_\mathrm{z}^{-1}, \label{eq:Menv_dot_ratio} \\
    \frac{M_\mathrm{env}}{M_\mathrm{env,0}} &= \frac{M_\mathrm{z} + M_\mathrm{xy} - M_\mathrm{c} }{M_\mathrm{z,0} + M_\mathrm{xy,0} - M_\mathrm{c,0}}
    \simeq 1+f_\mathrm{z}^{-1}\left(\epsilon-\epsilon_0\right), \\
    \frac{M_\mathrm{p}}{M_\mathrm{p,0}} &= \frac{M_\mathrm{z} + M_\mathrm{xy}}{M_\mathrm{z,0} + M_\mathrm{xy,0}} 
    \simeq 1+\epsilon-\epsilon_0. \label{eq:Mp_ratio}
\end{align}
\end{subequations}
The envelope's energy scales as $E_\mathrm{env} \propto M_\mathrm{env}^n$. If we combine this scaling with Eqs. \ref{eq:L_scaling_new} - \ref{eq:Mp_ratio} in normalized notation ($\tau_\mathrm{KH,0}=1$), this yields
\begin{equation}\label{eq:de_dtau_diffeq}
    \frac{d\epsilon}{d\tau} = n^{-1} f_\mathrm{z} {\bigg(\frac{\epsilon}{\epsilon_0}\bigg)}^{-\frac{\gamma_\mathrm{g}(1+\beta)}{\gamma_\mathrm{g}-1}}
    {\bigg(1 + \epsilon  - \epsilon_0} \bigg)^\frac{\gamma_\mathrm{g}+\beta}{\gamma_\mathrm{g}-1}
    {\bigg(1 + f_\mathrm{z}^{-1} (\epsilon-\epsilon_0)\bigg)}^{(1-n) + \frac{1+\beta}{\gamma_\mathrm{g}-1}}.
\end{equation}
Eq. \ref{eq:de_dtau_diffeq} has no insightful analytical solution, but can be evaluated iteratively. It can be reduced further in the heavily polluted limit where both $\epsilon \ll 1$ and $f_\mathrm{z}^{-1} \epsilon \ll 1$. This also leads to $n \approx 1$, and is equal to the limit that $(M_\mathrm{p,0} \gg M_\mathrm{xy,0}) \land (M_\mathrm{env,0} \gg M_\mathrm{xy,0})$ in Eq. \ref{eq:L_scaling_new}. This yields the simple ODE
\begin{equation}
    \frac{d\epsilon}{d\tau} \simeq f_\mathrm{z} {\left(\frac{\epsilon}{\epsilon_0}\right)}^{-\frac{\gamma_\mathrm{g}(1+\beta)}{\gamma_\mathrm{g}-1}}.
\end{equation}
Hence, the envelope's accretion follows the expression that we give in Eq. \ref{eq:cooling_polluted_lowmass}:
\begin{equation}
    \frac{\epsilon}{\epsilon_0} = \frac{M_\mathrm{xy}}{M_\mathrm{xy,0}} = {\left(1 + \frac{f_\mathrm{z}}{\epsilon_0} \frac{(2+\beta)\gamma_\mathrm{g}-1}{\gamma_\mathrm{g}-1} \frac{t}{\tau_\mathrm{KH,0}} \right)}^\frac{\gamma_\mathrm{g}-1}{(2+\beta)\gamma_\mathrm{g}-1}.
\end{equation}

}

\section{The fates of impactors}\label{Appendix_B}
\normalsize{There are several works that consider impact physics to predict the outcomes of accretion events, where the impactors collide with hot and dense proto-planetary envelopes (\citealt{Podolak1988, Mcauliffe2006, Mordasini2015, Pinhas2016}; \citetalias{Brouwers2018}; \citealt{Valletta2018}). The study by \citet{Mordasini2015} notably provides a comprehensive overview on the fates of differently sized impactors from pebbles to planetesimals. Their calculation is inspired by the Shoemaker-Levy 9 impact on Jupiter. Physically, it is based on a fragmentation model that assumes breakup to be a relatively slow process, driven by the occurrence of Rayleigh-Taylor instabilities. 

Most recent impact models approach fragmentation differently, that is as a rapid explosive event caused by the sudden disappearance of structural integrity when the dynamical pressure exceeds the planetesimal's internal strength. \citetalias{Brouwers2018} motivates this view analytically by estimating the time required to double the initial radius after breakup, which is only around $\sim$ 10 s. Here we supplement the impacts overview provided by \citet{Mordasini2015} with a similarly comprehensive figure, based on this explosive breakup. We designed a new code for this purpose, more simple than the one presented in \citetalias{Brouwers2018}. Instead of their fully self-consistent calculation, we used the analytical structure equations that we derived in the main text, where we assumed (pre-impact) grain-free and metal-free envelopes at 0.1 AU. 

Impactors are immediately affected by the increased gas drag when they enter an envelope. For pebbles, this means that their downward velocities soon approach the terminal velocity. Large planetesimals, by contrast, are mostly unaffected by the surrounding gas and stay close to the planet's local escape velocity. If impactors move too fast through the dense interior, the dynamical (ram) pressure that acts on their frontal surface can overcome the compressive strength that holds them together. In such an event, the impactor shatters and breaks up into a rapidly expanding cloud. The criterion for such a breakup is given by \citep[e.g.,][]{Valletta2018}
\begin{equation}
    \rho_\mathrm{g} v_\mathrm{i}^2 > S_\mathrm{i},
\end{equation}
where $R_\mathrm{i}$ is an impactor's radius and $S_\mathrm{i}$ is its compressive strength. The latter is a poorly constrained quantity and generally depends on the impactor's composition and formation history. Modeling of fragmentation events in the Earth's atmosphere and testing of the impactors post-impact yields compressive strengths in the broad range of 1-500 MPa \citep{Chyba1993, Svetsov1995, Petrovic2002, Popova2011, Podolak2015}, from which we take an approximate lower-range typical value $S_\mathrm{cst}$ of 5 MPa for a 100-m SiO$_2$ rock. We account for the fact that bigger impactors typically have greater effective strengths due to the importance of self-gravity, that we approximate with a scaling relation with exponent $\frac{3}{2}$ \citep{Benz1999, Steward2011}}:
\begin{equation}
 S_\mathrm{i} = S_\mathrm{cst} {\left(\frac{R_\mathrm{i}}{10^4 \; \mathrm{cm}}\right)}^\frac{3}{2}.
\end{equation}
\begin{figure}[t!] 
\centering
\includegraphics[width=\hsize]{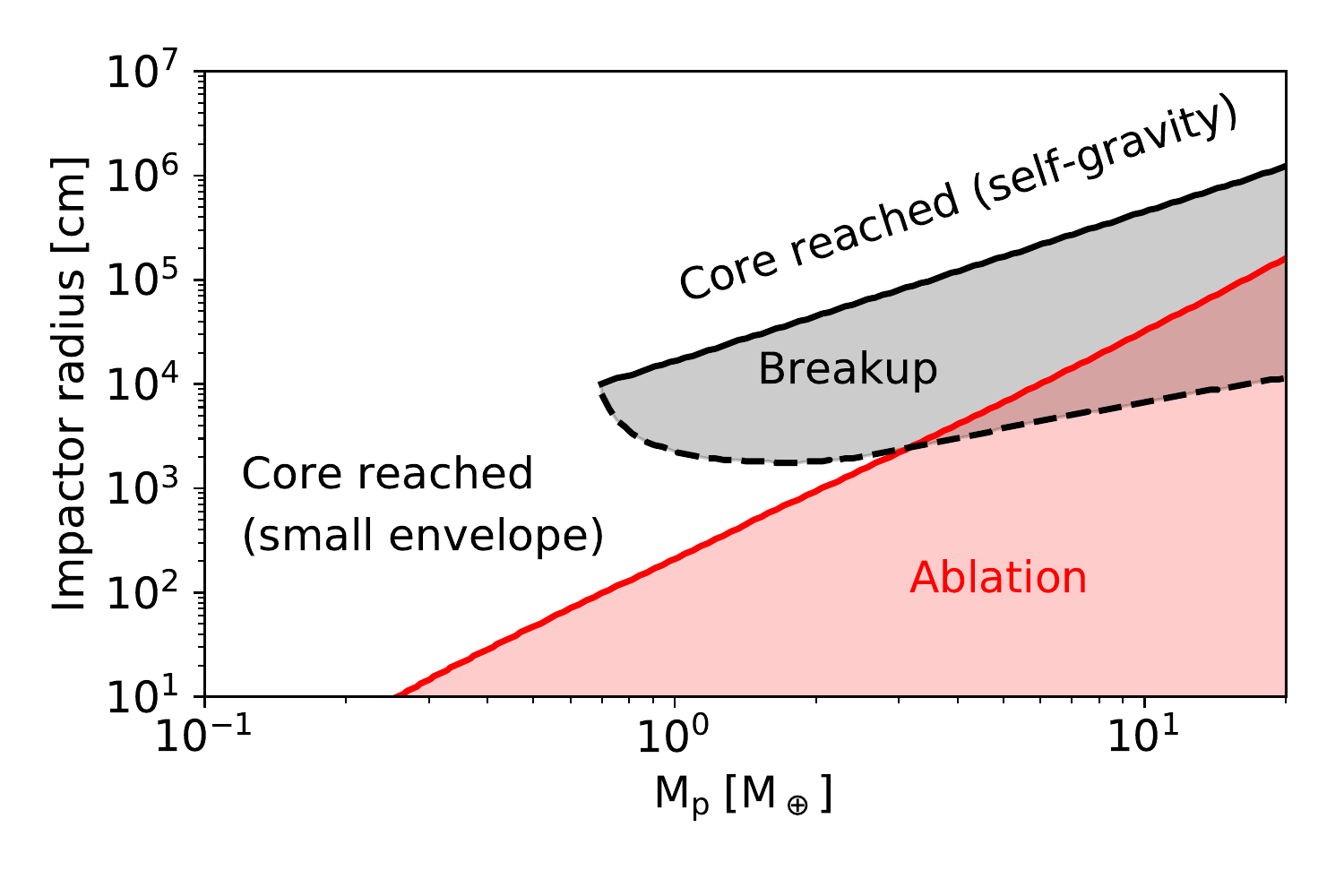} 
\caption{Fates of impactors during their accretion onto grain-free proto-planetary envelopes at 0.1 AU. The red zone indicates thermal ablation, the dominant disruption process for pebbles. It shows a proportional relationship with planetary mass, and agrees with our analytical expression of Eq. \ref{eq:R_crit_simple}. The grey zone indicates breakup. Small km-sized planetesimals are very susceptible to fragmentation, whereas 100 km planetesimals are almost completely resistant to it. \label{fig:impact_regimes}} 
\end{figure}

Aside from breakup, smaller impactors can also deposit their mass when they ablate in the envelope. We model this with simple energy-limited evaporation from the hot surrounding gas, as described by Eq. \ref{eq:P_in}. We do not consider frictional heating for two reasons: first, there is great uncertainty on the friction parameter. Including it therefore warrants a parameter study like \citet{Valletta2018} performed, instead of choosing a single value. Second, friction is expected to be most significant in regions where the ram pressure is high, the same regions that are already affected by fragmentation. Therefore, including it is not likely to contribute significantly to the final fate of impactors.

We use this simplified model to simulate impacts with objects between 10 cm and 100 km onto planets with masses in the range of 0.1 - 20 $\mathrm{M}_\oplus$ and plot the result in Fig. \ref{fig:impact_regimes}. We distinguish four regimes that describe the fates of impactors. Note that the location and size of these zones will shift depending on the assumed impactor composition, compressive strength, as well as the location in the disk and choice of opacity.
\begin{enumerate}[I]
    \item Impactors that accrete onto small planets ($\lesssim 0.3 \; \mathrm{M}_\oplus$) make it through the envelope and reach the core. These planets are too light to obtain sufficiently large envelopes to reach the internal temperatures necessary for SiO$_2$ evaporation. Their weak gravitational pull also means that the impactors travel too slowly to undergo breakup.
    \item{Pebbles that impact onto more massive planets vaporize in the hot interior and fail to reach the core. This is the main motivation for the work we present here.}
    \item{Small planetesimals ($\lesssim$ 1 km) are most susceptible to breakup and fail to reach the core unless the planet is very small. These impactors fall in between the two protected regimes of efficient drag and significant self-gravity.}
    \item{Large planetesimals ($\gtrsim$ 10 - 100 km) are strengthened by gravity and can reach the cores of more massive planets, depending on their size. Very large planetesimals, as are predicted to form through the streaming instability \citep{Johansen2007, Johansen2009, Simon2017, Liu2019, Li2019}, are effectively immune to fragmentation and can only be (partially) disrupted by efficient thermal ablation in the deep interior of gas giants.}
\end{enumerate}
\end{appendix}

\end{document}